\documentclass[iop,appendixfloats]{emulateapj}

\newbox\grsign \setbox\grsign=\hbox{$>$}
\newdimen\grdimen \grdimen=\ht\grsign
\newbox\laxbox \newbox\gaxbox
\setbox\gaxbox=\hbox{\raise.5ex\hbox{$>$}\llap
     {\lower.5ex\hbox{$\sim$}}}\ht1=\grdimen\dp1=0pt
\setbox\laxbox=\hbox{\raise.5ex\hbox{$<$}\llap
     {\lower.5ex\hbox{$\sim$}}}\ht2=\grdimen\dp2=0pt

\newcommand{\HAR}{rHARM}
\newcommand{\dF}{{^{^*}\!\!F}}

\usepackage[usenames,dvipsnames]{color}

\usepackage{CJKutf8}
\usepackage{amsmath}
\usepackage{ctable} 
\usepackage{graphicx}
\usepackage{natbib}
\usepackage[titletoc,title]{appendix}

\begin{document}
%\tracingall
\title{\MakeLowercase{r}HARM: ACCRETION AND EJECTION IN RESISTIVE GR-MHD}
\shorttitle{GR-MHD accretion-ejection}
\shortauthors{Qian et al.}
\author{
        Qian Qian (\begin{CJK*}{UTF8}{gbsn}{钱前\!\!}\end{CJK*})\altaffilmark{1},
        Christian Fendt\altaffilmark{1}, 
        Scott Noble\altaffilmark{2},
        Matteo Bugli\altaffilmark{3}
}
\altaffiltext{1}{Max Planck Institute for Astronomy, Heidelberg, Germany}
\altaffiltext{2}{Department of Physics and Engineering Physics, University of Tulsa, Tulsa, USA}
\altaffiltext{3}{Max Planck Institute for Astrophysics, Garching, Germany}
\email{ fendt@mpia.de, qian@mpia.de}                                   
% \date{\today}
%/////////////////////////////////////////////////////////////////////////
\begin{abstract}
Turbulent magnetic diffusivity plays an important role for accretion disks and the launching of disk winds.
We have implemented magnetic diffusivity, respective resistivity in the general relativistic MHD code HARM.
This paper describes the theoretical background of our implementation, its numerical realization, our 
numerical tests and preliminary applications.
The test simulations of the new code \HAR\ are compared with an analytic solution of the diffusion 
equation and a classical shock tube problem.
We have further investigated the evolution of the magneto-rotational instability (MRI) in tori around 
black holes for a range of magnetic diffusivities.
We find indication for a critical magnetic diffusivity (for our setup) beyond which no MRI develops 
in the linear regime and for which accretion of torus material to the black hole is delayed.
Preliminary simulations of magnetically diffusive thin accretion disks around Schwarzschild black holes 
that are threaded by a large-scale poloidal magnetic field show the launching of disk winds with mass fluxes 
of about 50\% of the accretion rate. 
The disk magnetic diffusivity allows for efficient disk accretion that replenishes the mass reservoir of 
the inner disk area and thus allows for long-term simulations of wind launching for more than 5000 time units. 
\end{abstract}

%//////////////////////////////////////////////////////////////////////////////////////
%////////////////////////////////////////////////////////////////////////////////////
%
\section{Introduction}
Relativistic jets as highly collimated beams of magnetized material moving with velocities close to the speed of light are 
found in various astrophysical objects, such as active galactic nuclei, X-ray binaries and most probably 
Gamma-ray burst sources \citep{1999ApJ...519L..17S,2001ApJS..134..181J,2006MNRAS.367.1432M,2006AIPC..856....1M}, 
while non-relativistic jets and outflows are ejected from young stellar objects \citep{2007prpl.conf..231R}.
A common understanding of all these jets sources is that they consist of a central object 
(e.g. a young star, neutron star or black hole) surrounded by an accretion disk which carries strong magnetic field.

In order to understand the origin of jets in such systems, the equations of magnetohydrodynamics (MHD) have to be 
solved. 
Obviously, for relativistic jets, launched from the close environment of a central black hole (BH), the treatment of 
general relativistic magnetohydrodynamics (in short resistive GR-MHD) is essential. 
Seminal papers suggest that jets can be disk driven, thus powered by the rotation of the inner accretion disk 
\citep{1982MNRAS.199..883B, 1986ApJ...301..571P, 2007prpl.conf..277P}.
In addition, for disks around rotating black holes, jet energy can be gained from the rotation of the black hole
magnetosphere \citep{1977MNRAS.179..433B}.

Given the fact that the physical equations governing such systems are non-linear, time-dependent, and intrinsically multidimensional, 
their solution is difficult and the link to the observational appearance is obstructed.
Furthermore, while jets from young stars can be rather well resolved and typical features such as mass fluxes, 
velocities, or even rotation can be observed \citep{2002ApJ...576..222B, 2004ApJ...604..758C}, 
the structure of the relativistic jet sources stays unresolved, and the dynamical jet parameters are uncertain 
as these jets are mostly detected in synchrotron emission. 

In the past, a few numerical schemes have been developed that are able to evolve the GR-MHD equations for the
process of relativistic jet formation. 
The first GR-MHD simulation was already presented by Wilson and Ruffini 
\citep{1975mnbh.book.....W,1975PhRvD..12.2959R, 1977mgm..conf..393W},
while it took another two decades before research returned to this topic.
The development of modern GR-MHD schemes has also been supported by the substantial advancement of CPU power. 
Milestones were the construction of non-relativistic MHD schemes (see e.g. \citealt{1992ApJS...80..753S}), 
special relativistic hydrodynamics codes \citep{2002A&A...390.1177D}, 
special relativistic MHD \citep{1999MNRAS.303..343K,2003A&A...400..397D}, 
general relativistic hydrodynamics  \citep{1984ApJ...277..296H,2002ApJ...577..866D},
and also MHD with pseudo-Newtonian potential \citep{2002ApJ...573..738H}. 

Today, quite a few GR-MHD codes have been accomplished 
\citep{1999ApJ...522..727K,2003ApJ...589..458D,2003ApJ...589..444G,2007A&A...473...11D},
being mainly used to simulate the evolution of black hole accretion systems 
(see e.g. \citealt{1999ApJ...522..727K,2003ApJ...592.1060D,2004ApJ...611..977M,2006MNRAS.368.1561M}).
All of the codes cited above work in the ideal MHD regime (thus neglecting resistivity or magnetic diffusivity). 
Steady-state GR-MHD accretion-outflow solutions were presented by \citet{2015ApJ...801...56P}.

In order to disentangle the powering mechanism for relativistic jets, one needs to investigate two important 
processes for jet formation, 
that is (i) the Blandford-Znajek effect, and (ii) the Blandford-Payne effect.
The Blandford-Znajek effect considers the fact that the rotational energy of a highly spinning black hole can be extracted
electro-magnetically by the magnetic field threading through the ergosphere \citep{1977MNRAS.179..433B}.
The Blandford-Payne effect considers a rotating disk magnetic field that allows - for a certain field inclination - to
accelerate material that is launched from the disk surface magneto-centrifugally along the field lines \citep{1982MNRAS.199..883B}. 
Ideal GR-MHD simulations of the Blandford-Znajek effect have been published \citep{2004ApJ...611..977M, 2006MNRAS.368.1561M, 2012MNRAS.423.3083M}.

The efficiency of jet formation from disks has been demonstrated by non-relativistic simulations \citep{1997ApJ...482..712O, 2002A&A...395.1045F} and
also special relativistic simulations \citep{2010ApJ...709.1100P, 2011ApJ...737...42P, 2013MNRAS.429.2482P}.
However the {\em launching} problem, the transition from accretion to ejection that requires the presence 
substantial amount of magnetic
diffusivity in order to allow for persistent disk outflows has not yet been treated in GR-MHD - presumably 
since resistive GR-MHD were not available.
The launching question is essential as it allows to compare the mass fluxes of disk and jet consistently, as it has been demonstrated by non-relativistic simulations of several groups
\citep{2002ApJ...581..988C, 2007A&A...469..811Z, 2012ApJ...757...65S, 2013ApJ...774...12F, 2014ApJ...793...31S, 2016arXiv160407313S}.

We believe that it is the absence of disk magnetic diffusivity in recent GR-MHD simulations that does not allow to form 
long-lasting powerful disk winds that may turn into jets.
With magnetic diffusivity a magnetized disk wind is launched and angular momentum of the orbiting disk material can 
be efficiently removed and allow for efficient accretion.

In the ideal GR-MHD regime, the matter cannot {"}cross{"} the magnetic field lines.
Such field lines that vertically thread the accretion disk are expected from advection of the magnetic flux from outer disk areas.
Thus, any mass that is ejected from the disk into the jet cannot be replenished by accretion from outer disk areas, and jet 
formation will decay.
In this case, the accretion flow will push the magnetic field lines inwards, which will accumulate out of the horizon of 
the BH and a magnetically arrested disk (MAD) will form that allows for further accretion only via the magnetic 
interchange instability \citep{2003ApJ...592.1042I,2003PASJ...55L..69N,2012MNRAS.423.3083M}.
Accretion via the interchange instability has also been found in non-relativistic 3D simulations of
protostellar disk around a dipolar stellar magnetosphere \citep{2012MNRAS.421...63R}.

In order to allow for a relative motion between plasma and magnetic field, we need to apply resistive MHD, 
thus magnetically diffusive MHD (see also \citealt{1997A&A...319..340F, 2002ApJ...581..988C}).
Besides being able to handle the re-distribution of mass flux that is needed for the launching mechanism of disk 
outflows, 
a resistive code can also treat physical magnetic reconnection \citep{2009ApJ...692..346F} that may explain the 
observed X-ray emission \citep{2003ApJ...585..429M}.

A number of resistive relativistic MHD codes have been developed, starting from \citet{2006ApJ...647L.123W} who 
investigated relativistic magnetic reconnection.
Pioneering work by \citet{2007MNRAS.382..995K} presented a multi-dimensional upwind scheme with resistivity in 
special relativity.
In \citet{2009MNRAS.394.1727P} an implicit-explicit (IMEX) Runge-Kutta method has been used to solve the stiff relaxation 
terms arising from resistivity.
That work has been further extended to three dimensions and general relativistic regime in \citet{2013PhRvD..88d4020D}. 
In \citet{2011ApJ...735..113T} a one-dimensional resistive approach has been undertaken in special relativistic 
regime using method of characteristics. 
% {\qq \citet{2012PhRvD..86d3002A} provides a complete model for resistive relativistic magnetohydrodynamics (this work 
% is not simulation!!)}. 
More recently, \citet{2013ApJS..205....7M} investigated the role of the equation of state in resistive GR-MHD. 

In order to develop our own resistive GR-MHD code\footnote{We denote our new code as \HAR}, we decided to follow the prescription of \citet{2013MNRAS.428...71B} who extended the 3+1 GR-MHD code ECHO \citep{2007A&A...473...11D} by 
implementing a mean-field dynamo closure and resistivity.

In the present paper we describe our implementation of resistivity respectively magnetic diffusivity, into the 
original ideal MHD code HARM.
We present various test simulations for our implementation.
We will further present astrophysical simulations comparing magnetically diffusive tori around rotating black holes with 
literature results obtained in ideal MHD, considering a possible decay of the magneto-rotational instability (MRI) in the resistive plasma.
We finally present preliminary simulations of thin disks around Schwarzschild black holes that are threaded by a large 
scale poloidal magnetic flux and launch outflows out of the accretion disk.

%=========================================================================================================
\section{Resistive GR-MHD equations}
\label{theo_eqs_chapter}
In the following we derive the equations of resistive MHD in general relativity that we have implemented 
in the existing ideal GR-MHD code HARM \citep{2003ApJ...589..444G, 2006ApJ...641..626N}.
In our derivation we follow \citet{2013MNRAS.428...71B}, who have also implemented resistivity in their code ECHO.
%We denote our version of the HARM code that includes resistivity with \HAR.
Significant changes to HARM were to be made, such as implementing new variables to describe the electric field and
the magnetic diffusivity.

We follow the conventional notation of \citet{1973grav.book.....M}, in particular the sign convention for the metric $(-,+,+,+)$.
Applying the Einstein summation convention, Greek letters have values $0,1,2,3$, while Latin letters take the 
values $1,2,3$.
The letter $t$ for indices denotes the zeroth component of a vector or a tensor. 
As in HARM, we apply the two observer frames that are defined by the co-moving observer, $u^{\mu}$, and the 
normal observer, $n^{\mu}$. 
The space-time of normal observer is split into the so called {"}3+1{"} form. 
The electric and the magnetic four vectors that are measured in the two frames are denoted by $e^{\mu}$, $b^{\mu}$ and 
$\mathcal{E}^{\mu}$, $\mathcal{B}^{\mu}$, respectively. 
For the normal observer frame we follow \citet{2006ApJ...641..626N} with the normal observer four velocity $n_{\mu}=(-\alpha,0,0,0)$ and the lapse time $\alpha = 1/\sqrt{-g^{tt}}$. 
Bold letters denote vectors while the corresponding thin letters with indices represent vector components.

As HARM, also \HAR\  is a conservative scheme, only that it evolves eleven {"}conserved{"} variables, instead of 
eight in HARM. 
Thus, eleven equations govern the time evolution of this set of variables. 
Correspondingly, these equations consider the conservation of mass, energy and momentum, and the evolution of the electric 
and the magnetic field.
Among these eleven equations only the equation for mass conservation can be taken from HARM,
\begin{equation} 
\frac{1}{\sqrt{-g}}\partial_{\mu}\left(\sqrt{-g}\rho u^{\mu}\right)=0
\label{particle_conservation}
\end{equation}
with $g \equiv det(g_{\mu \nu})$ and the mass density $\rho$. 
The equations considering the conservation of energy-momentum keep their general form,
\begin{equation} 
\partial_{t}\left(\sqrt{-g}T^{t}_{\,\,\mu}\right)+\partial_{i}\left(\sqrt{-g}T^{i}_{\,\,\mu}\right)
  = \sqrt{-g}T^{\kappa}_{\,\,\lambda}\Gamma^{\lambda}_{\,\,\mu \kappa}
\label{eq_ene-mom-cons}
\end{equation}
where $\Gamma^{\lambda}_{\,\,\nu \kappa}$ is the connection and $T^{\mu}_{\,\,\nu}$ is the stress-energy tensor 
consisting of a fluid part and an electromagnetic (EM) part,
\begin{equation} 
T^{\mu \nu} = T^{\mu \nu}_{\rm fluid} + T^{\mu \nu}_{\rm EM}.
\label{eq_stress_energy_tensor}
\end{equation}
Here, a difference to the ideal GR-MHD equations arises. 
The general definition for $T^{\mu \nu}_{\rm EM}$ is
\begin{equation} 
T^{\mu \nu}_{\rm EM} = F^{\mu \alpha}F^{\nu}_{\,\,\,\alpha} -\frac{1}{4} g^{\mu \nu} F_{\alpha \beta}F^{\alpha \beta}.
\label{eq_stress_energy_EM}
\end{equation}
For the ideal GR-MHD version of $T^{\mu \nu}_{\rm EM}$ we refer to e.g. \citet{2003ApJ...589..444G}.
For the resistive case, the $T^{\mu \nu}_{\rm EM}$ needs to consider the electric field, thus, to implement the 
anti-symmetric Faraday tensor,
\begin{eqnarray} 
F^{\mu \nu} & = & u^{\mu}e^{\nu} - e^{\mu}u^{\nu} + \epsilon^{\mu \nu \lambda \kappa} u_{\lambda} b_{\kappa}
\nonumber \\
F_{\mu \nu} & = & u_{\mu}e_{\nu} - e_{\mu}u_{\nu} + \epsilon_{\mu \nu \lambda \kappa} u^{\lambda} b^{\kappa} 
\label{eq:fara-tens}
\end{eqnarray}
in $T^{\mu \nu}_{\rm EM}$.
Here we use the Levi-Civita tensors 
\begin{equation}
\epsilon_{\alpha \beta \gamma \delta} =             \sqrt{-g}[\alpha \beta \gamma \delta], \quad
\epsilon^{\alpha \beta \gamma \delta} = - \frac{1}{\sqrt{-g}}[\alpha \beta \gamma \delta],
\end{equation}
with the conventional permutation symbol $[\alpha \beta \gamma \delta]$.
For $u^{\mu}$, we may use the four velocity of an arbitrary observer,
while $ e^{\mu}, b^{\mu}$, respectively, are the electric and the magnetic field measured in this certain frame. 
Similar equations hold for the dual Faraday tensor $^\ast\!F^{\mu\nu}$.
After some lengthy algebra , we have for the electromagnetic energy-momentum tensor 
\begin{eqnarray} 
T^{\mu \nu}_{EM} & = & \left(b^{2} + e^{2}\right) \left(u^{\mu}u^{\nu} + \frac{1}{2}g^{\mu \nu}\right) - b^{\mu} b^{\nu} - e^{\mu}e^{\nu} 
\nonumber \\
   & - & u_{\lambda}e_{\beta}b_{\kappa}\left(u^{\mu}\epsilon^{\nu \lambda \beta \kappa}
       + u^{\nu}\epsilon^{\mu \lambda \beta \kappa}\right),
\label{eq_stre_ene_EM_2}
\end{eqnarray}
which is in agreement with \citet{2006MNRAS.367.1797M}. 
In order to avoid confusion, we point out that the sign convention in the definition of the Levi-Civita tensors in this paper 
follows \citet{1973grav.book.....M}, which differs from the convention used in \citet{2013MNRAS.428...71B}. 
Eventually, the stress-energy tensor that we apply in \HAR\ becomes 
\begin{eqnarray} 
T^{\mu \nu} 
    & = & \left(\rho + u + p+ b^{2} + e^{2}\right) u^{\mu}u^{\nu} + \left(p + \frac{1}{2}\left(b^{2} + e^{2}\right)\right) g^{\mu \nu} 
\nonumber \\
    & - & b^{\mu} b^{\nu} - e^{\mu}e^{\nu} - u_{\lambda}e_{\beta}b_{\kappa}
\left(  u^{\mu}\epsilon^{\nu \lambda \beta \kappa} + u^{\nu}\epsilon^{\mu \lambda \beta \kappa}\right).
\label{eq_str_ene_tens_2}
\end{eqnarray}
Here, $u$ is the internal energy, $p$ denotes the gas pressure and  $b^{2}=b^{\mu}b_{\mu}$, $e^{2}=e^{\mu}e_{\mu}$. 

In \HAR\ both the electric and the magnetic field are evolved in the normal observer frame. 
The evolution of the magnetic field four vector follows from the Maxwell equations,
\begin{equation} 
\partial_{t}(\sqrt{-g}\, \dF^{it}) = -\partial_{i} \left( \sqrt{-g}\, \dF^{ij} \right),
\label{B_evolution}
\end{equation}
and the constraint 
\begin{equation} 
\partial_{i}(\sqrt{-g}\, \dF^{it}) = 0.
\label{B_constraint}
\end{equation}
Similar to the modification of the stress-energy tensor, also the complete form of dual Faraday tensor,
\begin{equation} 
\dF^{\mu \nu} = -u^{\mu}b^{\nu} + b^{\mu}u^{\nu} - \epsilon^{\mu \nu \lambda \kappa} u_{\lambda} %e_{\kappa},
\label{duel_Faraday_tensor}
\end{equation}
is required for the EM field evolution here.

With the definition 
$\mathcal{B}^{i} \equiv n_{\nu}\,\dF^{\nu i} = \alpha \dF^{i t}$ and $\mathcal{B}^{0} \equiv n_{\nu} \dF^{\nu t} = 0$, 
a direct relation follows between the $\dF^{it}$ and the magnetic field four vector $\mathcal{B}^{\mu}$, 
that is the magnetic field in normal observer's frame. 
Similarly, the electric field four vector is defined by $\mathcal{E}^{i} \equiv n_{\nu}F^{i \nu}=-\alpha F^{it}$, while $\mathcal{E}^{0}=0$. 
In order to present the equations more comprehensively, we will use $\mathcal{B}^{\mu}$, $\mathcal{E}^{\mu}$ for the 
theoretical derivation instead of $\dF^{it}$, $-F^{it}$, which are actually used in \HAR.   

The time evolution of the electric field four vector $\mathcal{E}^{\mu}$ follows from 
\begin{eqnarray} 
\gamma^{-1/2} \partial_{t} \left(\gamma^{1/2}\pmb{\mathcal{E}}\right) 
    & - & \nabla \times \left(\alpha \pmb{\mathcal{B}} - \pmb{\beta} \times \pmb{\mathcal{E}} \right)
          + \left(\alpha \pmb{v} - \pmb{\beta} \right) \nonumber \\
    & = &  -\alpha \Gamma \left[ \pmb{\mathcal{E}}  +\pmb{v} \times \pmb{\mathcal{B}} 
          - \left(\pmb{\mathcal{E}} \cdot \pmb{v}\right) \pmb{v} \right] / \eta,
\label{eq_E_evol}
\end{eqnarray}
\citep{2013MNRAS.428...71B},
where $\pmb{\beta}=\{\beta^{i}\}$ is the spatial shift vector in
3+1 formalism, $\Gamma$ denotes the Lorentz factor and $\gamma = \sqrt{-g}/\alpha$ is the determinant of its spatial 3-metric.
The $\pmb{v}$ denotes the three velocity in the normal observer frame (see Section \ref{2D+1_Scheme_subsec}).
Here a new variable enters the system of equations, namely the resistivity or magnetic diffusivity $\eta$ 
(see next section for details).
Note that equation (\ref{eq_E_evol}) is the combination of the two Maxwell equations for the electric field
\begin{eqnarray} 
\gamma^{-1/2} \partial_{t} \left(\gamma^{1/2}\pmb{\mathcal{E}}\right) 
    - \nabla \times \left(\alpha \pmb{\mathcal{B}} - \pmb{\beta} \times \pmb{\mathcal{E}} \right) & = & 
    - \left(\alpha \pmb{\mathcal{J}} - q\pmb{\beta} \right), 
\nonumber \\
    \nabla \cdot \pmb{\mathcal{E}} & = & q
\label{Maxwell_eq_E}
\end{eqnarray}
and the resistive condition $e^{\mu} = \eta j^{\mu}$ (see also section 3), where $q$ is the electric 
charge density, $\pmb{\mathcal{J}}$ is the electric current for the normal observer and $j^{\mu}$ denotes 
the components of the electric current in the co-moving observer frame (see \citealt{2013MNRAS.428...71B} 
for a detailed derivation).

%====================================
\section{Resistivity in \HAR}
\label{res_in_rHARM_chatper}
Essentially, two new physical quantities enter the system of equations in \HAR. 
These are the electric field variable and the resistivity (or magnetic diffusivity) $\eta = \eta (r,\theta)$.

We understand the resistivity as due to turbulence, thus closely related to the alpha-viscosity in
turbulent accretion disks \citep{1973SvA....16..756S, 1973A&A....24..337S}.
In the following we briefly motivate the use of a turbulent resistivity, mainly quoting from derivations 
presented by \citet{1996A&A...307..665K} and \citet{2013MNRAS.428...71B}.

Starting from classical Ohm's law $\pmb{E}+\pmb{v}\times \pmb{B} = \eta_{\rm o} \pmb{J}$ 
with the (microscopic) resistivity $\eta_{\rm o}$
and assuming turbulent fluctuations in the velocity $\pmb{v}'$, the electric field $\pmb{E}'$, and the
magnetic field $\pmb{B}'$,
mean-field averaging will lead to a revised mean electric current
\begin{equation} 
\overline{\pmb{E}}+\overline{\pmb{v}}\times \overline{\pmb{B}} =
- \overline{\pmb{v}' \times \pmb{B}'} + \eta_{\rm o} \overline{\pmb{J}},
\label{classical_Ohms_law_2}
\end{equation}
with the mean-field velocity and fields  $\overline{\pmb{v}}, \overline{\pmb{B}}, \overline{\pmb{E}}$, respectively.
The term $\overline{\pmb{v}' \times \pmb{B}'}$ does not vanish, since the fluctuating quantities inside are
presumably correlated. 
A key assumption is that this term can usually be written as a linear combination of both the mean magnetic field and 
its curl, namely
\begin{equation}
\overline{\pmb{v}' \times \pmb{B}'} = - \alpha_{\rm D} \overline{\pmb{B}} - \eta_{\rm t} \nabla \times \overline{\pmb{B}},
\label{classical_Ohms_law_3}
\end{equation}
where the two scalars $\alpha_{\rm D}$ and $\eta_{\rm t}$ are isotropic coefficients and are both proportional to 
the local turbulent correlation time. 
With this assumption and dropping the bars from now on, we can rewrite Ohm's law equation as 
\begin{equation} 
\pmb{E}+\pmb{v}\times \pmb{B} = \alpha_{\rm D} \pmb{B} + (\eta_{\rm t} + \eta_{\rm o}) \pmb{J}.
\label{classical_Ohms_law_4}
\end{equation}
The $\alpha_{\rm D}$-term may introduce exponentially growing modes and is usually known as mean-field dynamo, 
while the $\eta_{\rm t}$-term acts as a resistivity - a turbulent resistivity. 
In this paper, we will not further consider the dynamo term (see \citealt{2014IJMPS..2860203B} for an application). 
Since we will mostly focus on the diffuse effect of resistivity, we also refer to it as magnetic diffusivity.
In astrophysical plasma, usually $\eta_{\rm t} \gg  \eta_{\rm o}$ and we will therefore 
apply $\eta \equiv \eta_{\rm t} \simeq \eta_{\rm t} + \eta_{\rm o}$ and write Ohm's law as
$\pmb{E}+\pmb{v}\times \pmb{B} = \eta \pmb{J}$, or, in covariant form of equation in the co-moving frame, where $v=0$,
\begin{equation}
e^{\mu} = \eta j^{\mu}.
\label{covariant_Ohm's_law_1}
\end{equation}
The scalar $\eta$ in equation (\ref{covariant_Ohm's_law_1}) is exactly the input diffusivity in the present version 
of \HAR. 
The ideal GR-MHD regime can be retrieved by setting $\eta=0$, namely $e^{\mu}=0$. 
Applying magnetic diffusivity, the numerical time stepping $\delta t$ requires to consider the diffusive time scale
on the grid scale, thus $\delta T_{\eta} < \Delta x^{2}/\eta$, where $\Delta x$ is the smallest cell size.
This turned out to be critical, especially when running the code with at relative large $\eta$ where the dynamic time 
scale becomes larger then the diffusive time scale.

%=====================================================================================================
\section{Inversion scheme}
\label{inversion_scheme_chapter}

\HAR\ uses the following set of {"}conserved{"} variables
\begin{equation} 
\pmb{U} \equiv \sqrt{-g}(D, T^{t}_{\,\,\,t}, T^{t}_{\,\,\,i}, ^\ast F^{it}, -F^{it}),
\label{conserved_variables}
\end{equation}
where $D \equiv \rho u^{t}$. 
The time evolution of $\pmb{U}$ is performed by using the equations derived in the last section. 
These equations are written in the so-called conserved form, for which the time derivative of the variable depends on
the position derivative of its {"}flux{"}. 
To model these fluxes $\pmb{F}$ for $\pmb{U}$ across the surfaces of the simulation cells, an additional set of so-called 
{"}primitive{"} variables is needed. 
Similar to the ideal HARM, the {"}primitive{"} variables in \HAR\ are
\begin{equation} 
\pmb{P} = (\rho,u,v^{i}, ^\ast\!F^{it},-F^{it})
\label{primitive_variables}
\end{equation}
where $\rho$ stands for density, $u$ for internal energy, and $v^{i}$ for the spatial 3-velocity for the normal observer.  
The $\dF^{it}$ and $-F^{it}$ are related to the magnetic and the electric four vectors for the normal observer by a factor of $\alpha$
(see Section \ref{theo_eqs_chapter}) and they are both conserved as well as primitive variables.

As discussed in \citet{2003ApJ...589..444G}, the variables $\pmb{U}(\pmb{P})$ and $\pmb{F}(\pmb{P})$ can be expressed as analytic 
functions of primitive variables, but the inverse operations do not have a closed-form. 
Hence, the numerical inversion scheme to extract $\pmb{P}$ from $\pmb{U}$ at each time step after the evolution of $\pmb{U}$
is the core of a conservative GR-MHD code. 

The resistive term in equation (\ref{eq_E_evol}) could become {\em stiff} \citep{2013MNRAS.428...71B}, as usually we deal with 
a small resistivity $\eta \lesssim 10^{-2}$. 
Unfortunately, the stiff term also contains $\pmb{\mathcal{E}}$, which makes $\pmb{\mathcal{E}}$ impossible to evolve
in time simultaneously with other conserved variables.
Therefore, its solution has to be found by some {\em implicit} scheme, e.g. together with primitive variables, and the 
inversion scheme used in ideal GR-MHD HARM must be emended in \HAR\ under the resistive context.

%===============================================================================================
\subsection{2D+1 Scheme}
\label{2D+1_Scheme_subsec}
Following \citet{2006ApJ...641..626N}, it is convenient to project the energy-momentum density into normal observer frame
\begin{equation} 
Q^{\mu} \equiv  -n_{\nu} T^{\mu \nu},
\label{Q_def}
\end{equation}
together with the projection tensors $j_{\mu \nu} = g_{\mu \nu} + n_{\mu}n_{\nu}$ , the energy-momentum flux perpendicular to the normal 
observer can be described by $\tilde{Q}^{\mu} = j^{\mu}_{\,\,\nu}Q^{\nu}$. 
${\tilde{Q}}^{\mu}$ is a four vector with its zeroth component always being zero and we note it as $\tilde{Q}^{i}$. 
Also we define $U \equiv -Q_{\mu}n^{\mu} = \alpha^{2}T^{00}$. 
Obviously, $\tilde{Q}^{i}$ and $U$ inherit the information from conserved variables $T^{t}_{\,\,\,\mu}$.

With the Lorentz factor $\Gamma$ we define $W \equiv (\rho + p + u)\Gamma^{2}$. 
The flow velocity relative to the normal observer is denoted by 
${\tilde{v}}^{i}=j^{i}_{\mu}u^{\mu}$ and 
$v_{i}={\tilde{v}}^{i}/\Gamma$, $v^{2} \equiv v_{i}v^{i}$. 
$\Gamma$ is a function of $v^{2}$, $\Gamma^{2}=1/(1-v^{2})$. 
The gas pressure $p$ is a function of $v^{2}$ and $W$ depends on the equation of state, which is $u=p/(\gamma_{\rm gas}-1)$ in \HAR. 
%{\qq note this pressure depends on the EOS!!!}

The variables to be solved by the inversion scheme are $\rho$, $v^{i}$ and  $\mathcal{E}^{i}$). 
The relation between the conserved variables and $\rho$, $v^{i}$ is given (with help of the above definitions) by 
\begin{eqnarray} 
D & = & \rho \Gamma \nonumber \\
\pmb{\tilde{Q}} & = & W  \pmb{v} + \pmb{\mathcal{E}} \times \pmb{\mathcal{B}} \nonumber \\	
              U & = & W - p + (\pmb{\mathcal{E}}^2 + \pmb{\mathcal{B}}^2)/2
\label{U(P)}
\end{eqnarray}
\citep{2006ApJ...641..626N, 2007A&A...473...11D}.
The ideal GR-MHD regime is retrieved by replacing the vector $\pmb{\mathcal{E}}$ with $-v \times \pmb{\mathcal{B}}$. 
However, in the resistive case, $\pmb{\mathcal{E}}$ needs to be solved separately.
Rewriting equation (\ref{eq_E_evol}) into the numerical form, the relation between $\mathcal{E}^{i}$ and other 
variables is \citep{2013MNRAS.428...71B} 
\begin{eqnarray} 
   \mathcal{E}^{i}& = &\{\epsilon^{ijk}\tilde{v}_{j}\mathcal{B}_{k} \nonumber \\
   	              & + &\tilde{\eta}[N^{i}+(N^{k}\tilde{v}_{k}\tilde{v}^{i})/(1+\tilde{\eta}\Gamma)]\}/(\Gamma+\tilde{\eta}),
\label{E_evolution_numerical_1}
\end{eqnarray}
where 
\begin{eqnarray} 
N^{i} & = & \mathcal{E}^{i(0)}  + \Delta t [ -\left(\alpha v^{i}-\beta^{i} \right) \gamma^{-1/2}\partial_{k} \left(\gamma^{1/2}\mathcal{E}^{k(0)}\right)   
\nonumber \\
& - &  \epsilon^{ijk}\partial_{j} \left( \alpha \mathcal{B}_{k}-\epsilon_{klm} \beta^{l} \mathcal{E}^{m} \right) ],
\nonumber \\
1 / \tilde{\eta} & = & \Delta \alpha / \eta .
\label{E_evolution_numerical_2}
\end{eqnarray}
The $N^{i}$ comes from the none {\em stiff} term which does not include $\eta$, hence can be solved explicitly. 
The $\mathcal{E}^{i(0)}$ and $\mathcal{E}^{k(0)}$ denote the electric field four vector from the last time step,
respectively. 
The sign flip before $\epsilon^{ijk}$ in comparison to \citet{2013MNRAS.428...71B} is due to the different definitions of Levi-Civita tensor mentioned in
section 2.

\citet{2006ApJ...641..626N} have suggested a way to combine equation (\ref{U(P)})
(under the condition $\pmb{\mathcal{E}}$ = $-v \times \pmb{\mathcal{B}}$) into an equation system with two equations 
only of conserved variables, $W$ and $v^{2}$. 
This equation system can eventually be solved by 2-dimensional Newton-Raphson (NR) method. 
We keep this feature in our inversion scheme and combined equation (\ref{U(P)}) by calculating 
$\pmb{\tilde{Q}}^{2}=\tilde{Q}^{i} \tilde{Q}_{i}$
\begin{eqnarray} 
\pmb{\tilde{Q}}^{2} & = & (W  \pmb{v} + \pmb{\mathcal{E}} \times \pmb{\mathcal{B}})^{2}
\nonumber \\
      		 & = & W^{2}\pmb{v}^{2} + (\pmb{\mathcal{E}} \times \pmb{\mathcal{B}})^{2} 
      		   - 2 W \pmb{\mathcal{E}} \cdot (v \times \pmb{\mathcal{B}}).
\label{Q^2_1}
\end{eqnarray}
To eliminate $\pmb{v}$, it is useful to calculate  
\begin{equation} 
\pmb{\tilde{Q}} \times \pmb{\mathcal{B}} =  W v \times \pmb{\mathcal{B}} 
        + (\pmb{\mathcal{E}} \times \pmb{\mathcal{B}}) \times \pmb{\mathcal{B}},
\label{QtcrossB}
\end{equation}
which gives the relation
\begin{equation} 
v \times \pmb{\mathcal{B}} = W^{-1}[\pmb{\tilde{Q}} \times \pmb{\mathcal{B}} 
                            - (\pmb{\mathcal{E}}  \times \pmb{\mathcal{B}} ) \times \pmb{\mathcal{B}} ].
\label{vcrossB}
\end{equation}
Inserting this into equation (\ref{Q^2_1}), with some simple algebra we obtain
\begin{equation} 
\pmb{\tilde{Q}}^{2} = v^{2}  W^{2} - (\pmb{\mathcal{E}} \times \pmb{\mathcal{B}})^{2} 
                                   - 2\pmb{\mathcal{E}} \cdot (\pmb{\tilde{Q}} \times \pmb{\mathcal{B}}).
\label{Q^2_2}
\end{equation}

Since $p$ is a function of $v^{2}$ and $W$, the last equation in equation (\ref{U(P)}) already satisfy the requirement and 
together with equation (\ref{Q^2_2}), they give an equation system only consists of conserved variables, $v^{2}$ and $W$
\begin{eqnarray} 
\pmb{\tilde{Q}}^{2}  -  \pmb{v^{2}}  W^{2} - (\pmb{\mathcal{E}} \times \pmb{\mathcal{B}})^{2} 
                     - 2\pmb{\mathcal{E}} \cdot (\pmb{\tilde{Q}} \times \pmb{\mathcal{B}}) &=& 0,   \nonumber \\	
         U - W + p(\pmb{v^{2}},W) - \frac{1}{2}(\pmb{\mathcal{E}}^2 + \pmb{\mathcal{B}}^2) &=& 0.
\label{NR_eqations}
\end{eqnarray}
For a given $\pmb{\mathcal{E}}$, Equation~\ref{NR_eqations} can be solved by a 2D NR-method. 
Once $v^{2}$ and $W$ are solved, $\rho$, $u$ and $v^{i}$ can be retrieved by
\begin{eqnarray} 
\rho & = & D(1-\pmb{v}^{2})^{1/2}
\nonumber \\
   v^{i} & = & W^{-1}[\tilde{Q}^{i} - {\pmb{\mathcal{E}} \times \pmb{\mathcal{B}}}^{i}],
\nonumber \\
   u & = & p/(\gamma_{\rm gas} -1).
\label{P(U)}
\end{eqnarray}

Nevertheless, $\pmb{\mathcal{E}}$ does not evolve with other conserved variables and cannot be considered as $given$ at the 
beginning of the inversion scheme. 
We solve this problem by considering an extra loop, which specifically makes $\pmb{\mathcal{E}}$ converge. 
In total, the inversion scheme to extract the primitive variables from the conserved variables in \HAR\ follows the steps as below.
\begin{enumerate}
  \item[1.] Take the conserved variables after a new time evolution, except the $\pmb{\mathcal{E}}$ that is taken
           from the former time step(or initial time step).
  \item[2.] Apply them to the two equations in Equation~\ref{NR_eqations} and solve for the primitive variables $u$, $v^{i}$ with the
           2D NR scheme.
  \item[3.] Renew $\pmb{\mathcal{E}}$ with the solution obtained in step 2 using equation (\ref{E_evolution_numerical_1}).
  \item[4.] Repeat step 2 and step 3 until $W$, $v^{2}$ and $\pmb{\mathcal{E}}$ converge.  
  \item[5.] Calculate the primitive variables using equation (\ref{P(U)}). 
\end{enumerate} 

Since this inversion scheme uses a 2D Newton-Raphson method with an additional extra loop over $\pmb{\mathcal{E}}$, we find
it convenient to denote it by the term {\em 2D+1 scheme}.  
Note that in ideal HARM, a series of inversion schemes have been included (see \citealt{2006ApJ...641..626N}), 
from which we yet only revised the 2D scheme and no other inversion scheme works in \HAR\ for now.
A schematic flow chart with a description of the numerical procedures for the time evolution in \HAR\ is shown in 
Appendix \ref{num_pro_chapter}.

\subsection{A preliminary test of the implementation}
The implementation of the electric field can be tested by running \HAR\ with $\eta \rightarrow 0$ (we used $\eta=10^{-12}$), comparing the evolution 
of the conservative variables and 
the fluxes from all grid cells generated by the primitive variables for a few time steps. 
We find that they coincide within errors of $10^{-10}$.
%They are practically equal with 2-10 equal digits after decimal mark in scientific notation to those from ideal HARM. 
We then compared the primitive variables obtained by the inversion scheme of the first time step with those obtained by the ideal HARM,
finding similar accuracy. 
We may thus confirm the correct implementation of the new stress-energy tensor and the new inversion scheme that 
now also considers the electric field. 
Further tests of \HAR\ considering $\eta >0$ will be discussed in Section \ref{diff_box_chapter} and Appendix \ref{shock_tube_test}.

%=====================================================================================================
\section{Model set up in the science simulations}
\label{model_setup_chapter}
In this section we briefly mention the common setup for \HAR, that are used for the numerical simulations 
presented later. 
We basically follow the setup in ideal HARM \citep{2003ApJ...589..444G,2006ApJ...641..626N} with minor additions in 
the boundary conditions for the electric field.

\subsection{Units and normalization}
The units we used throughout in the simulations in this paper have $GM=c=1$, which sets the length unit $r_{\rm g} \equiv GM/c^{2}$ 
and time unit $t_{\rm g} \equiv GM/c^{3}$. The black hole angular momentum $J=jGM^{2}/c$, $j=a/M$ is the dimensionless Kerr 
parameter with $-1 \leqslant j \leqslant 1$. We use $r_{\rm H} = 1 + \sqrt{1-j^{2}}$ to denote the event horizon, which varies with the black hole angular momentum. The densities and mass fluxes in this paper are presented in the code unit.

\subsection{Numerical grid}
\label{num_grid}
The numerical integrations are carried out on a uniform grid with a so-called modified Kerr-Schild coordinates: 
$x_{0}$, $x_{1}$, $x_{2}$, $x_{3}$, where $x_{0}=t$, $x_{3}=\phi$ are the same as in Kerr-Schild coordinates, 
while the radial, and $\theta$ coordinates are calculated by the relation:
\begin{eqnarray} 
r &=& R_{0}+e^{x_1},
\nonumber \\
\theta &=& \pi x_2 + \frac{1}{2} (1-h) \sin(2\pi x_2).
\label{MKS_r_theta}
\end{eqnarray}
Different $R_{0}$ and $h \in [0,1]$ will return different concentration of grid resolution in radial and $\theta$ direction. 
A smaller $h$ value indicates a better concentration in $\theta$ around equatorial plane.
Except for the 1D shock tube test problem presented in Appendix \ref{shock_tube_test} that employs a flat space-time 
with uniform Cartesian coordinates, we used Kerr-Schild coordinates with $R_0 = 0$ for all simulations presented
in this paper.
Note that for scalars and the $r$ and $\theta$ vector components this presentation is invariant to Boyer-Lindquist 
coordinates, only the time component and the $\phi$-component transform, so the time slicing is different.

\subsection{Boundary conditions and the initial condition for electric field}
We apply outflow condition at inner and outer boundary, for which all the primitive variables are projected into the 
ghost zones while forbidding inflow at inner and outer boundary.
Both axial boundaries have reflection condition, where the primitive variables are projected in to the ghost zone 
with a mirror effect.
Boundary conditions for the electric field have been added for \HAR, similar to those for the magnetic field\footnote{Note that for our test simulations of \HAR, a variety of geometrical setups and boundary conditions are used (see Section \ref{diff_box_chapter} and Appendix \ref{num_pro_chapter} for detail)}. 
The initial electric field is chosen to be equal the ideal MHD value, $\pmb{\mathcal{E}}=-\pmb{v} \times \pmb{\mathcal{B}}$. 
It turned out that this choice works very well in the non-ideal MHD simulations.

%-----------------------------------------------------------------------------------------------------
\section{Test simulations of magnetic diffusivity}
\label{diff_box_chapter}
In this section, we present a test for the implementation of magnetic diffusivity by comparison with an analytic 
solution of the diffusion equation.
Further test problems are discussed in the Appendix.
Her we follow the procedure suggested by \citet{2002A&A...395.1045F}. 
A similar approach was presented by \citet{2013MNRAS.428...71B}, the evolution of a self-similar current sheet.
Our test simulations are performed in a small, almost rectangular box of hydrostatic gas at varying distance from the black hole. 
The gas in the box is {"}heavy{"} and is penetrated by a {"}weak{"} magnetic field - such that dynamical effects due to 
Lorentz forces are negligible and the magnetic field distribution changes only by diffusion.
We have applied different levels of diffusivity (set constant in the domain).
We compare the results to the known analytic solution and find a perfect match between the numerical and the analytical 
results.

\begin{figure*}
\centering
\includegraphics[width=3.5in]{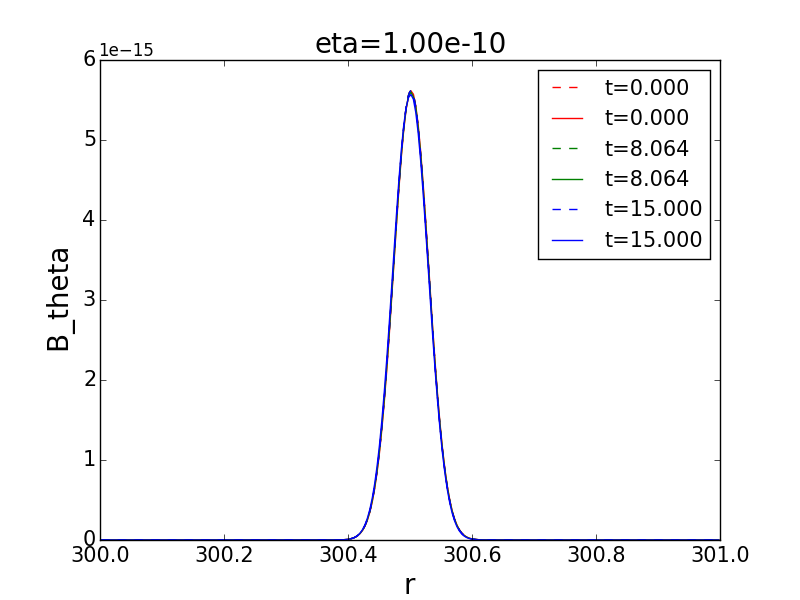}
\includegraphics[width=3.5in]{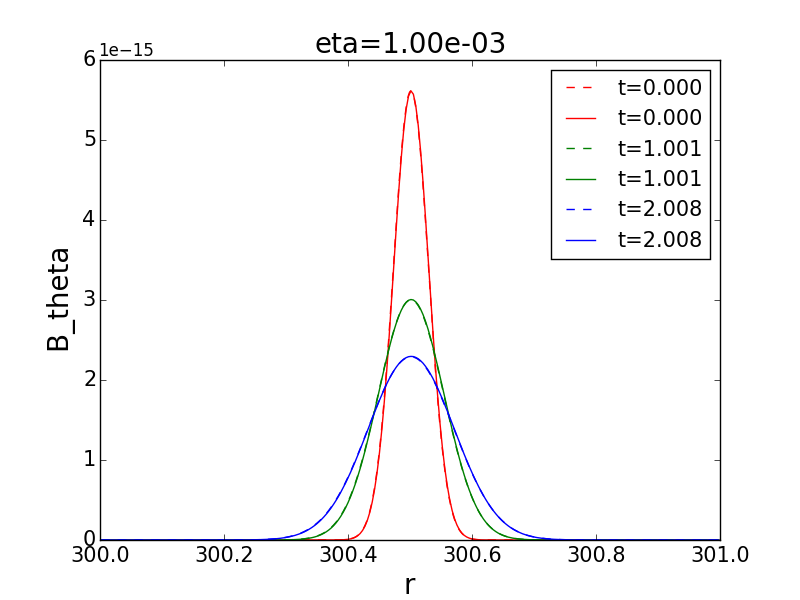}
\includegraphics[width=3.5in]{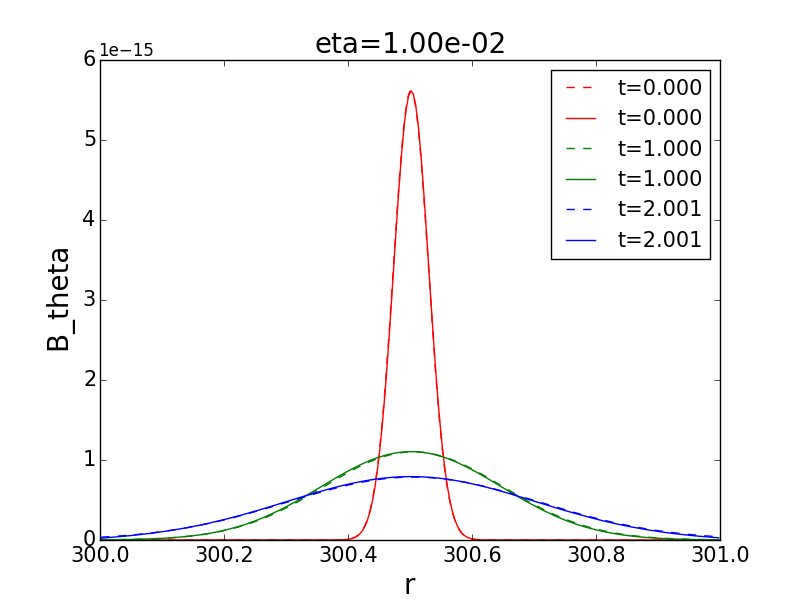}
\includegraphics[width=3.5in]{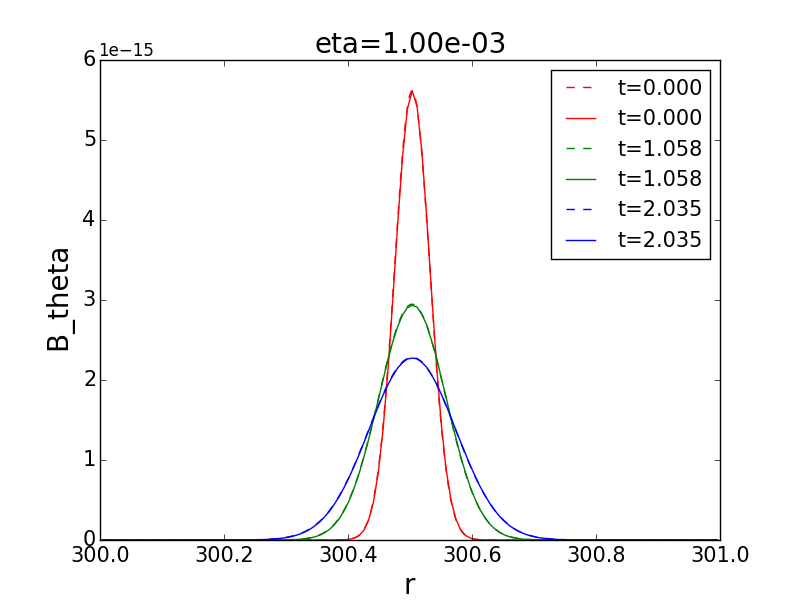}
\caption{Numerical tests of magnetic diffusivity. 
The radial profile of the magnetic field $B_{\theta}(r)$ is plotted along the equatorial plane. 
We show simulation {\em difT0} with $\eta = 10^{-10}$ (upper left),
simulation {\em difT1} with $\eta = 10^{-3}$ (upper right),
simulation {\em difT2} with $\eta = 10^{-2}$ (lower left),
all with a grid resolution 256x256,
and simulation {\em difT3} with $\eta = 10^{-3}$
and a grid resolution 128x128 (lower right).
Different colors represent the corresponding simulation time steps $t$ in the legend. 
The actual time $\tilde{t}$ of these steps are $\tilde{t} = t_{0}+t$ where $t_{0}$ depends on the initial condition. 
Solid lines are from simulation results while dashed lines from analytic solutions.
Note the difficulty in distinguishing dashed lines from solid lines, since 
the perfect match between the analytical and the numerical evolution.
In the upper left plot, all 6 curves are actually plotted. 
Still, they look like one curve, since with $\eta=10^{-10}$ the magnetic field does not diffuse at all.}
\label{res_test_300to301}
\end{figure*}

%------------------------------------------------------------------------
\subsection{Simulation region and boundary condition}
The simulation area is chosen as a small sub-sector of the axisymmetric spherical uniform grid along the equatorial plane,
and small enough that the shape of the area is rectangular to a high degree.
The size of this area extends $\Delta R$ in radius and $\Delta \theta $ in latitude, and is located at 
a radius $r_0$.
The concentration parameter is set to $h=1$, which means the spacing in $\theta$ direction is uniform.
For $\Delta r \ll r_{0}$ the sector can indeed be treated as a rectangular box with
$r\in [r_{0}-\Delta r/2, r_{0}+\Delta r/2]$, $\theta \in [\pi/2-\Delta \theta/2, \pi/2+\Delta \theta /2]$ 
and $\Delta r = r_{0} \Delta \theta = R $.

In the test simulations the Kerr parameter $a=0$ and the event horizon in this case is the Schwarzschild radius, $r_{H}=2$. 
A continuous outflow boundary condition is set for all the four boundaries of the box. 
Our simulations are denoted by {\em difT0}, {\em difT1}, {\em difT2}, {\em difT3}, and {\em difT4} 
(see Table~\ref{box_table}). 

%----------------------------------------------------------------------------------
\subsection{Initial conditions}
We apply a relativistic gas with polytropic index $\gamma_{\rm G}=4/3$. 
Initially, the gas in the box is in hydrostatic equilibrium.
Both the density profile and the magnetic field profile are set to be uniform in $\theta$ direction. 
In radial direction, the density profile was set such that the pressure gradient cancels the gravity,
\begin{equation} 
   \nabla_{r}p(r) = -\rho(r) r^{-2}.
\label{pressure_gradient}
\end{equation}
For simplicity, we have used a Newtonian potential in equation (\ref{pressure_gradient}). 
This choice works well for large distances from the black hole.
Close to the black hole, our choice is inconsistent with GR.
In particular, a hydrostatic state does not exist anymore (see below).
Nevertheless, these inconsistencies are small and did not really influence our main conclusion concerning
the test of magnetic diffusivity. 
From equation (\ref{pressure_gradient}) we apply the following radial profiles for density and pressure in
the box,
\begin{eqnarray} 
   \rho (r) & = & C \cdot r^{\alpha},
\nonumber \\
       p(r) & = & \beta \cdot \rho^{\gamma_{\rm G}},     
\label{density_pressure_profile}
\end{eqnarray}
where $\alpha=1/(1-\gamma_{\rm G})$, $\beta=1/(1-\alpha)$ and $C$ denotes a proper normalization constant. 

For the magnetic field, we only consider the $\theta$ component.
The initial field strength $B_{\theta}$ is chosen from the solution of the one dimensional diffusion equation for 
infinite space resembling a Gaussian profile with time evolution  
\begin{equation}
B_{\theta}(r,\tilde{t})=\frac{1}{\sqrt{\tilde{t}}} \exp\left(-\frac{(r-r_{0})^{2}}{4\eta \tilde{t}}\right).
\label{B_theta_ana}
\end{equation}
Here, 
$\tilde{t}=t_{0}+t$ and in our context $t$ is the actual code running time.
The parameter $t_{0}$ then is defined by the choice of the peak value of the initial Gaussian profile. 
The initial $B_{\theta}$ is thus defined by $t=0$.
The solution equation (\ref{B_theta_ana}) will be later be compared to our simulation results.
We choose a very weak magnetic field with a plasma beta $\beta \equiv p_{\rm gas}/p_{\rm mag}=10^{8}$. 
The diffusivity $\eta$ is set to be uniform throughout the simulation region. 

\begin{table}
\caption{Test simulations of magnetic diffusivity $\eta$ in \HAR.
Parameter choice in simulations {\em difT0}, {\em difT1}, {\em difT2}, {\em difT3}, {\em difT4}.
The table shows the radial position of the domain in the equatorial plane and the domain size,
both in units of $r_{\rm g}$,
and the grid resolution in the domain.
The magnetic diffusivity $\eta$ is constant in the domain.
Simulations were done for $a=0$}
\begin{center}
%  \begin{tabular}{ c | c | c | c | c }
   \begin{tabular}{ c c c c c }
    \hline  \noalign{\smallskip}  
      run  & $\eta$     & distance from origin & domain size & grid size \\ 
    \noalign{\smallskip}   \hline
    \hline  \noalign{\smallskip}  
  {\em difT0}  & $10^{-10}$ & 300$r_{\rm g}$     & 1x1$r_{\rm g}^{2}$      & 256x256 \\
  {\em difT1}  & $10^{-3}$  & 300$r_{\rm g}$     & 1x1$r_{\rm g}^{2}$      & 256x256 \\
  {\em difT2}  & $10^{-2}$  & 300$r_{\rm g}$     & 1x1$r_{\rm g}^{2}$      & 256x256 \\
  {\em difT3}  & $10^{-3}$  & 300$r_{\rm g}$     & 1x1$r_{\rm g}^{2}$      & 128x128 \\
  {\em difT4}  & $10^{-3}$  &  30$r_{\rm g}$     & 5x5$r_{\rm g}^{2}$      & 256x256 \\
    \noalign{\smallskip}   \hline
  \end{tabular}
  \end{center}
\label{box_table}
\end{table}

\begin{figure*}
\centering
\includegraphics[width=3.5in]{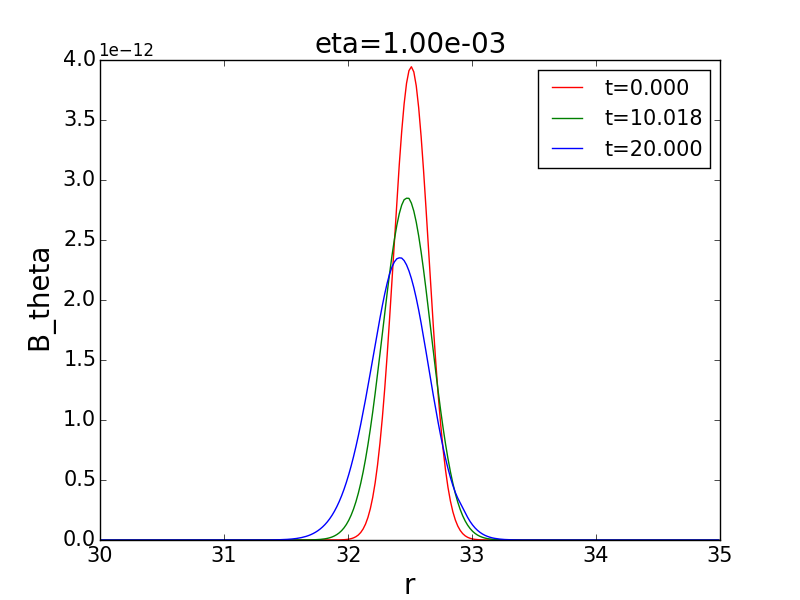}
\includegraphics[width=3.5in]{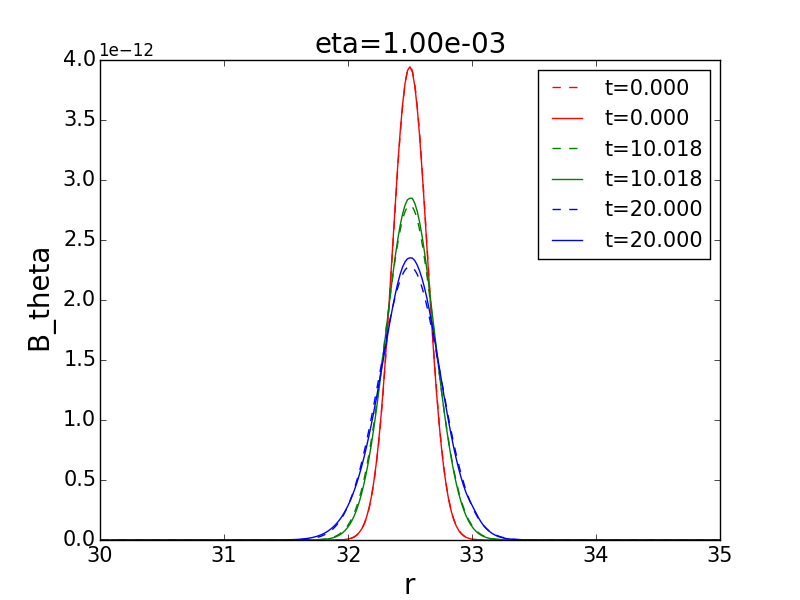}
\caption{
Shown is simulation {\em difT4} with $\eta = 10^{-3}$ in a box of 5x5 $r_{\rm H}$ located at $r = 30 r_{\rm H}$
and with a grid resolution 256x256.
The radial profile of the magnetic field $B_{\theta}(r)$ is plotted along the equatorial plane. 
The left plot shows advection of magnetic flux by the infalling corona.
The right plot shows the same simulation results, but compensated for advection / infall with
the magnetic profile maximum shifted back to the center of the simulation box (after the simulation).
In the right plot, solid lines are from the simulation result while dashed lines from the analytic solution. 
}
\label{res_test_30to35}
\end{figure*}

%=====================================================================================
\subsection{Test simulations of magnetic diffusivity}
In this subsection, we show the results of tests simulation {\em difT0} to {\em difT3}
with a box of size of 1x1 $r_{\rm g}$ being placed at $r_{0}=300$.
At this distance, GR effects can be neglected and the box can be safely be considered as {"}rectangular{"}. 
The grid resolution is 256x256 in general except for simulation {\em difT3} with a grid size of 128x128 in order
in the interest of exploring resolution effects.

In order to compare the simulation results to the analytic solutions, we show the magnetic field $B_{\theta}(r)$ 
along the equatorial plane $\theta=\pi/2$ at different time steps (see Figure \ref{res_test_300to301}).
Numerical results are plotted in solid lines while the analytic solution is shown in dashed lines.

In the first test, we ran simulation {\em difT0} considering a tiny resistivity $\eta=10^{-10}$ (upper left).
As a result, we retrieve the behavior of the ideal MHD gas such that the magnetic field did not diffuse at all.
Simulations {\em difT1} and {\em difT2} apply a high diffusivity with $\eta=10^{-3}$ and $\eta=10^{-2}$, respectively
(upper right and lower left).
In both cases the initial Gaussian profiles decay nicely following exactly analytic solution. 
Simulation {\em difT3} with two times lower resolution (lower right) performed similarly well, 
except the fact that the peak of the Gaussian is not as well resolved as before.

%--------------------------------------------------------------------------------
\subsection{Diffusivity test near a black hole}
It is essential to test the implementation of magnetic diffusivity in \HAR\ also for regions closer 
to the black hole.
In order to test the code performance in this regime, we have set up simulation run {\em difT4}, for
which we choose a {"}box{"} size of 5x5 and a box location located at $r_{0}=30 r_{
\rm g}$.
We show the results in Figure \ref{res_test_30to35}.

The left plot shows that besides the diffusive decay of the magnetic field, the magnetic flux is also advected inwards.
The velocity of this motion is about $v_r \simeq - 10^{-3}$ at time $t=3 t_{
\rm g}$.
This effect can also be observed in the simulations discussed previously, only that the radial velocities 
are much lower (about $10^{-5}$ at $t=3 t_{\rm g}$), and thus negligible. 
In the simulation runs discussed in Appendix \ref{num_diff_chapter}, the run time is comparatively much longer,
so that the acceleration towards the black hole can be seen more clearly.
We attribute this effect to our choice of an initially hydrostatic corona, derived using a Newtonian potential
(Thus, this setup becomes more inconsistent with GR for small radii). 
The gas in the computational domain will start to fall towards the black hole, and will thereby advect magnetic flux.
We can, however, easily disentangle this effect and compare the magnetic diffusion in the simulation 
with the analytic solution.
In order to do so, we have shifted the profile of the magnetic field distribution resulting from the simulations 
outwards to the center of the simulation area, compensating for the advection of magnetic flux.
We can see (right plot) that the shifted curves and the analytic solution are in very good agreement,
again approving our implementation of magnetic diffusivity.

We note that the numerical curves are slightly higher compared to the analytic solutions in this case. 
This can be understood by the in-fall of gas that comes along also with a compression of gas and magnetic field. 
Close to the black hole the simulation area is rather a sector of a ring than a rectangular box
(note that $r_{0} \Delta \theta = 5 r_{\rm g} \sim r_{0}$).

%================================================================================
\section{Simulations of resistive magnetized tori}
\label{sect_tori}
In this section, we apply our resistive GR-MHD code to a problem that is astrophysically more relevant - the 
evolution of a magnetized torus near a rotating black hole. 
We follow the general setup in \citet{2003ApJ...589..444G} prescribing an axisymmetric torus of rotating gas 
around with a magnetic field distribution that is confined in the torus. 

We will present simulations with two choices of grid resolution and also different values for a magnetic diffusivity that is 
constant in space and time. 
For comparison, another simulation is executed using the original ideal HARM code \citep{2006ApJ...641..626N}.
The parameters of our simulation runs are listed in Table \ref{torus_table}.
Simulations denoted by {\em torT0...2} apply a grid resolution of 256x256 and are intended to further test the 
implementation of magnetic diffusivity in \HAR\ by comparing {\em torT1} and {\em torT2} to the ideal MHD HARM 
simulation {\em torT0}. 
Simulations denoted by {\em mriT1...6} apply a grid resolution of 128x128 and intend to survey how magnetic diffusivity 
affects the evolution of the magneto-rotational instability (MRI) in the torus.  

\begin{figure*}
\centering
\includegraphics[width=3.5in]{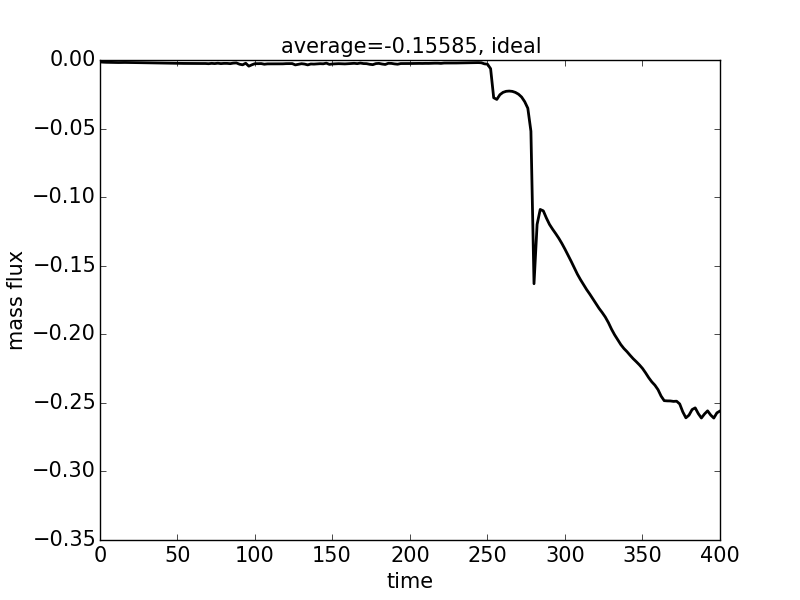}
\includegraphics[width=3.5in]{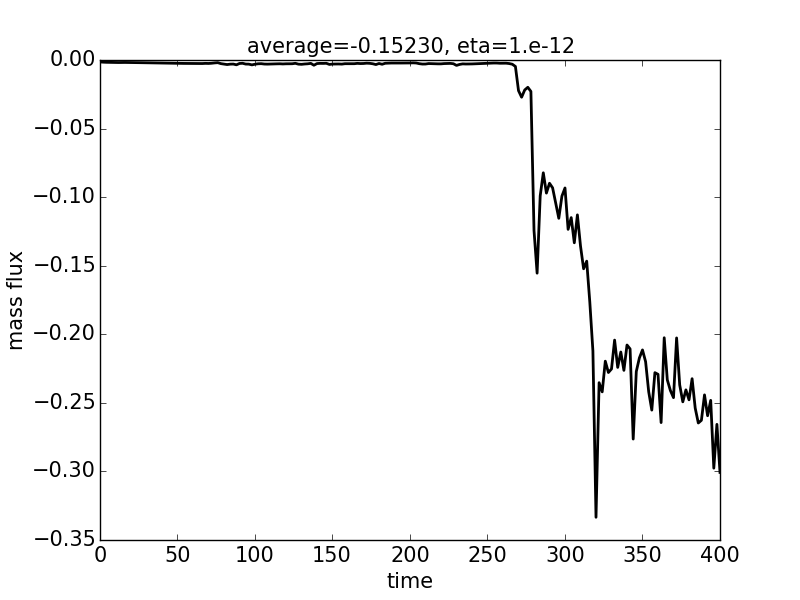}
\includegraphics[width=3.5in]{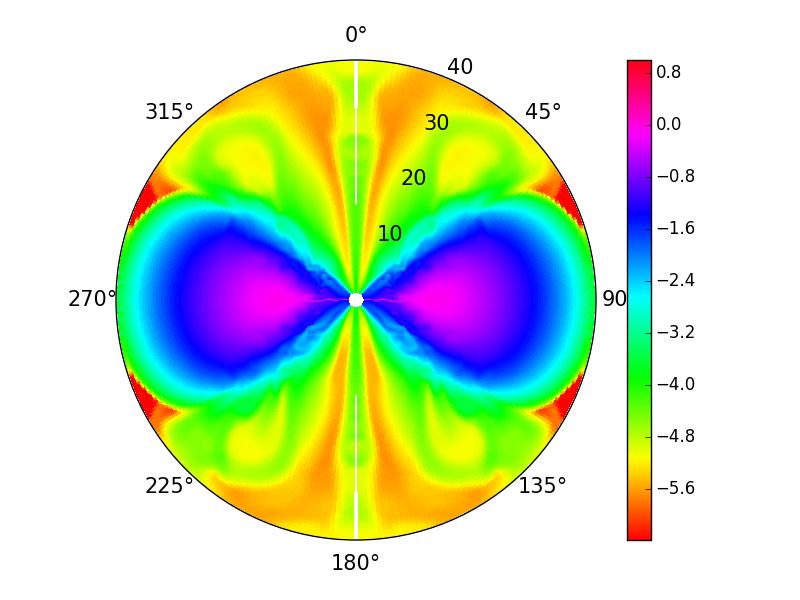}
\includegraphics[width=3.5in]{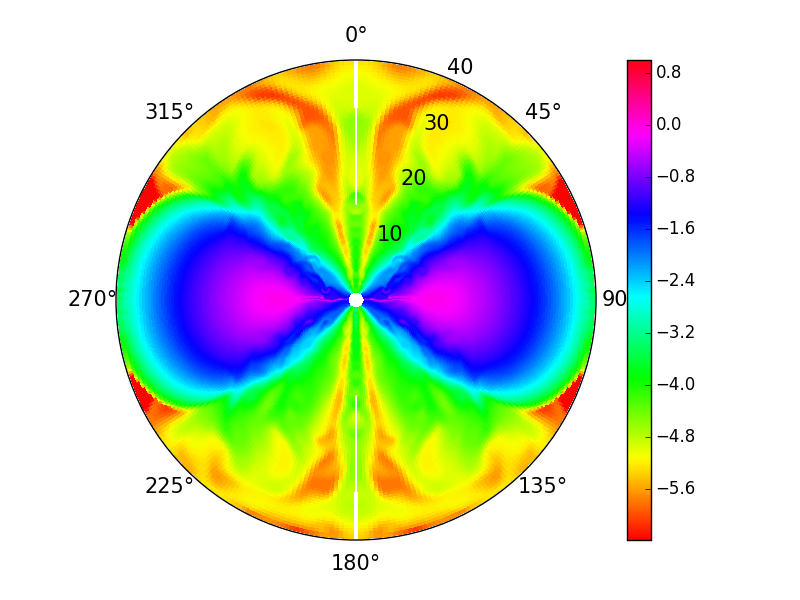}
\caption{The figure shows the comparison of simulations {\em torT0} and {\em torT1}.
Simulation {\em torT0} applies ideal MHD HARM  while {\em torT1} applies \HAR\ with tiny magnetic diffusivity 
$\eta = 10^{-12}$. 
Mass accretion rates in simulation {\em torT0} (upper left) and {\em torT1} (upper right) 
are measured at $r=2.2r_{\rm g}$ close to the horizon. 
The average accretion rates shown in the plot title are taken from $t=240t_{\rm g}$ to $t=400t_{\rm g}$. 
Density ($log(\rho)$) distribution of simulation {\em torT0} (lower left) and {\em torT1} (lower right) 
at $t=400t_{\rm g}$ (the domain covers only the right side of the plot, the left part is just mirrored). 
}
\label{torus_1}
\end{figure*}

\subsection{Computational domain and initial conditions}
%
% The numerical computation of the simulations is done in an axisymmetric domain. 
The computational domain is an axisymmetric half sphere with the radius ranging from $r_{\rm in}=0.98 r_{\rm H}$ 
to $r_{\rm out}=40r_{\rm g}$.
For all the simulations, 
the angle $\theta$ ranges from $0$ to $\pi$ and the grid concentration parameter is set to $h=0.3$.
As we apply a Kerr black hole with $a/M=0.9375$, the event horizon $r_{\rm H} \approx 1.35t_{\rm g}$ (see Section \ref{model_setup_chapter}).

The simulations in this section evolve an equilibrium gas torus surrounding a black hole, which is a particular 
solution of the class of equilibria found by \citet{1976ApJ...207..962F} and \citet{1978A&A....63..221A}. 
The torus is embedded in a vacuum (of a certain floor density).
Centrifugal forces and gas pressure in the torus balance gravity (see also \citealt{2003ApJ...589..444G}).

The torus inner edge is set at $(r,\theta)=(6,\pi/2)$ and the pressure maximum is located at $r_{\rm max}=12 r_{\rm g}$.
The orbital period of the torus at the radius of pressure maximum is about $267t_{\rm g}$, measured by an observer 
at infinity. 
A polytropic equation of state $p = (\gamma_{\rm G} -1)u$ is applied with $\gamma_{\rm G} =4/3$. 

The purely poloidal initial magnetic field consists of concentric field lines superposed on the density
contours of the equilibrium torus applying a vector potential $A_{\phi} \propto {\rm max}(\rho/\rho_{\rm max}-0.2,0)$
\citep{2003ApJ...589..444G}.
The field is normalized such that the minimum value of the plasma beta is 
$\beta =p_{\rm gas}/p_{\rm mag}=10^{2}$ \citep{2003ApJ...589..444G}.

The diffusivity in the torus simulations is constant in space and time. 
According to our previous test results of \HAR, a diffusivity $\eta = 10^{-12}$ in simulation {\em torT1} will
retrieve the ideal MHD regime of HARM, 
while with $\eta = 10^{-3}$, diffusive MHD effects should appear in simulation {\em torT2}.
For the MRI simulations, we choose a range of diffusivity - between $\eta = 10^{-12}$ and $\eta = 10^{-3}$ - in order
to scan the impact of diffusivity on the evolution of the MRI.
%The set up of simulation T0 is the same as other simulations, except that it does not have the diffusivity input, since 
%ideal HARM only evolves the ideal GR-MHD equations. 
%{\rot 
%The final time of the simulations is chosen to $t=2000$.
%}

\begin{table}[h]
\caption{Parameter choice in the torus simulations that use \HAR\ and ideal HARM. The table shows the value of diffusivities, innermost boundaries, resolutions that are used in the torus simulations. Except that {\em torT0} was executed by ideal HARM, all other tests were done by {\HAR}. The Kerr parameter is $a=0.9375$ in all simulations, thus $0.98 r_{\rm H} = 1.32 r_{\rm g}$. }
\begin{center}
%  \begin{tabular}{ c | c | c | c }
  \begin{tabular}{ c c c c c }
    \hline   \noalign{\smallskip}  
         & $\eta$          & grid size    & code \\ 
     \noalign{\smallskip}   \hline
    \hline  \noalign{\smallskip}  
    {\em torT0}  & -                  & 256x256         & HARM\\
    {\em torT1}  & $10^{-12}$         & 256x256         & \HAR  \\
    {\em torT2}  & $10^{-3}$          & 256x256         & \HAR  \\
    {\em mriT1}  & $10^{-12}$         & 128x128         & \HAR  \\
    {\em mriT2}  & $10^{-6}$          & 128x128         & \HAR  \\
    {\em mriT3}  & $10^{-4}$          & 128x128         & \HAR  \\
    {\em mriT4}  & $5\times10^{-4}$   & 128x128         & \HAR  \\
    {\em mriT5}  & $8\times10^{-4}$   & 128x128         & \HAR \\
    {\em mriT6}  & $10^{-3}$          & 128x128         & \HAR \\
     \noalign{\smallskip}   \hline
  \end{tabular}
  \end{center}
\label{torus_table}
\end{table}

\subsection{Robustness of \HAR\ as seen from torus simulations}
For most of the simulations presented in this paper the inversion scheme converged to high accuracy
for almost all grid cells.
However, under certain conditions - such as very low magnetic diffusivity or a very strong magnetic 
field - convergence might fail. 
While the strong-field limit is a typical problem of MHD codes in general, the case of low resistivity 
will usually not be applied with a resistive code (we applied this only for testing the implementation 
of resistivity).
Still, to advance the inversion scheme for applications of the code in these regimes will be one of the next
steps in developing our code further.

Due to the problems mentioned just above we cannot compare the diffusive simulations {\em torT1} and 
{\em torT2} over the same period of time as the ideal MHD simulation {\em torT0} was running.

For example, simulation {\em torT1} is supposed to retrieve the evolution of {\em torT0} for which massive accretion
of matter sets in as soon as the MRI is established.
However, even in the in the ideal HARM simulation, we observe that the inversion scheme is {"}overburdened{"}
at cells close to the horizon when a density discontinuity develops after massive accretion starts.
The scheme returns primitive variables with less accuracy and even fails to converge on singular grid cells 
after time $t=250t_{\rm g}$, which is about the time of first accretion impact (see Figure \ref{torus_1}). 
This problem is augmented in \HAR\ for the sake of the extra loop to make electric field variables converge in 
the 2D+1 scheme.
In the end this somewhat diminishes the robustness of the inversion scheme in the present version of \HAR.

\begin{figure*}
\centering
\includegraphics[width=3.5in]{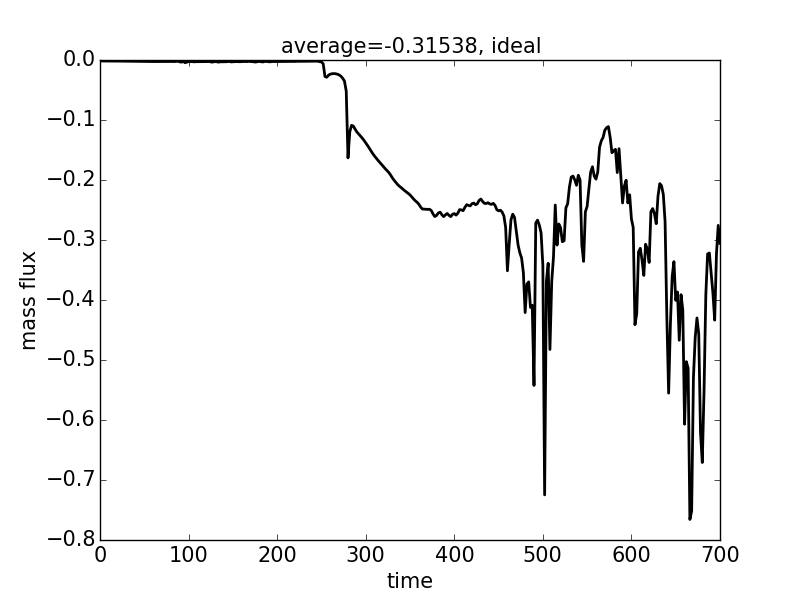}
\includegraphics[width=3.5in]{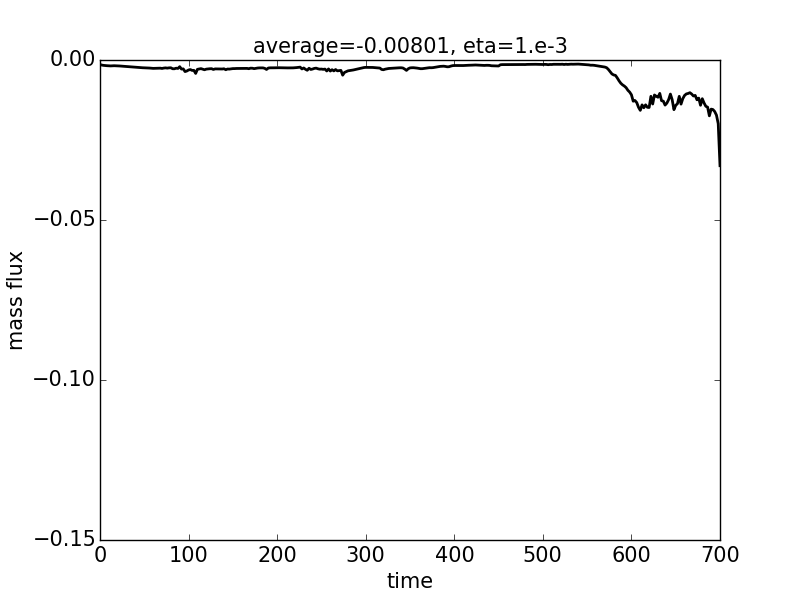}
\caption{Mass accretion rates of simulation {\em torT0} (left plot) and {\em torT2} (right plot) at $r=2.2r_{\rm g}$. 
The left plot in Figure \ref{torus_1} is actually a part of this plot. 
The averages values show in the plot titles were taken from $t=500t_{\rm g}$ to $t=700t_{\rm g}$. 
A continuous accretion appeared in {\em torT0} after about $t=300t_{\rm g}$, while no massive accretion observed 
in {\em torT2} during the simulation.}
\label{torus_3}
\end{figure*}

Note that also the simulation {\em torT2} runs longer than simulation {\em torT1} since the accretion rate in {\em torT2}
is much lower (and hence produces a milder density jump).
Still, we can compare simulation {\em torT0} to the data from simulation {\em torT1, torT2} for a limited period
of evolution.
In the following, we compare results of these simulations.

\subsection{Comparing simulation {\em torT1} and {\em torT2} to simulation {\em torT0}}
We first compare the results from simulation {\em torT0} and {\em torT1}. 
As mentioned above, these two simulations are supposed to be consistent with each other. 
We calculate their mass accretion rates at $r=2.2r_{\rm g}$ from $t=0$ to $t=400t_{\rm g}$ and plot them in 
Figure \ref{torus_1} (upper plots). 
The accretion rates($\dot{M}$) are calculated using
\begin{eqnarray} 
\dot{M}(r) = \int_{0}^{\pi} 2 \pi \rho(r,\theta) u^{r}(r,\theta) \sqrt{-g} d\theta. 
\label{mass_flux_def}
\end{eqnarray}
As can be seen from the plots for the accretion rate, the tori in both simulations keep their equilibrium state until 
the angular momentum transport supported by MRI \citep{1991ApJ...376..214B,2003ApJ...589..444G} finally results in 
accretion after at about $t=220t_{\rm g}$. 
The features of accretion beginning in simulation {\em torT0} are retrieved quite well in simulation {\em torT1}. 
Figure \ref{torus_1} (lower plots) show the density plots at $t=400t_{\rm g}$ of the two simulations where the 
accretion already starts a while and begins to disturb the surface of the gas torus. 
Basically, the $\dot{M}$ calculation and the density plot from simulation {\em torT1} nicely match those of  
simulation {\em torT0}. 

\begin{figure*}
\centering
\includegraphics[width=3.5in]{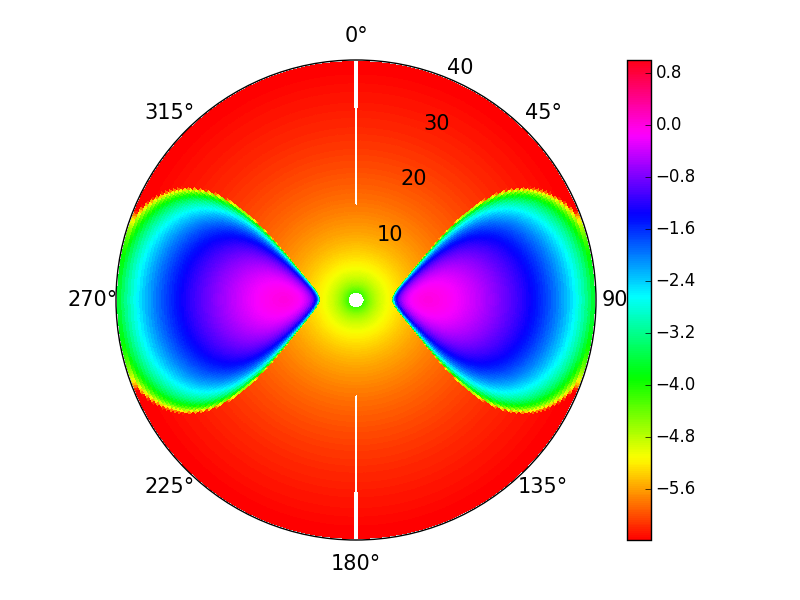}
\includegraphics[width=3.5in]{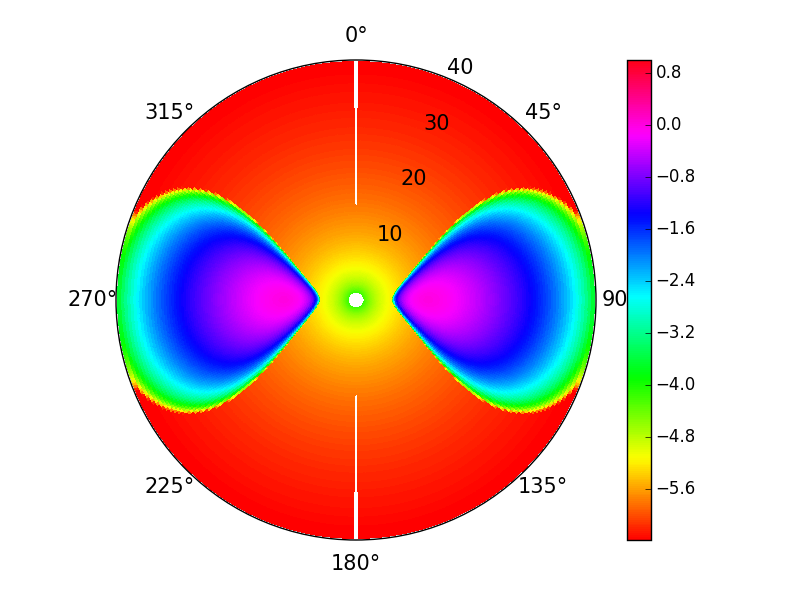}
\includegraphics[width=3.5in]{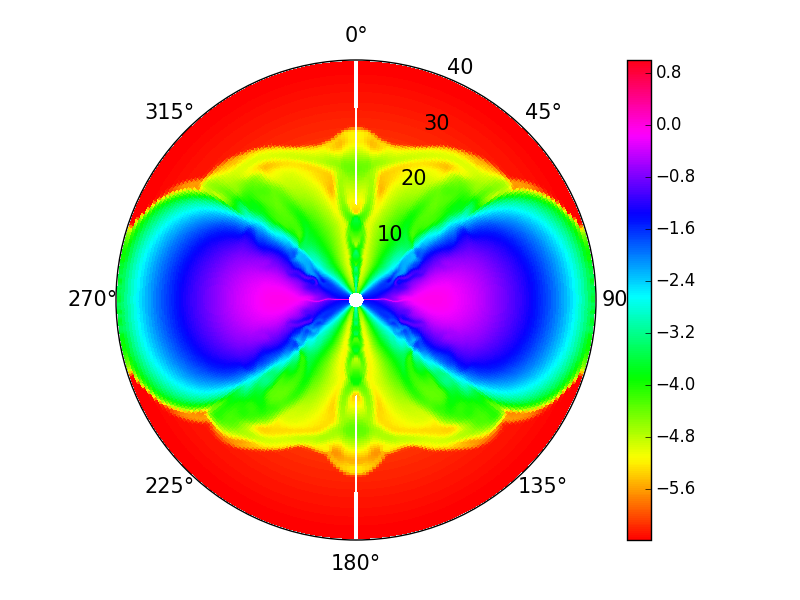}
\includegraphics[width=3.5in]{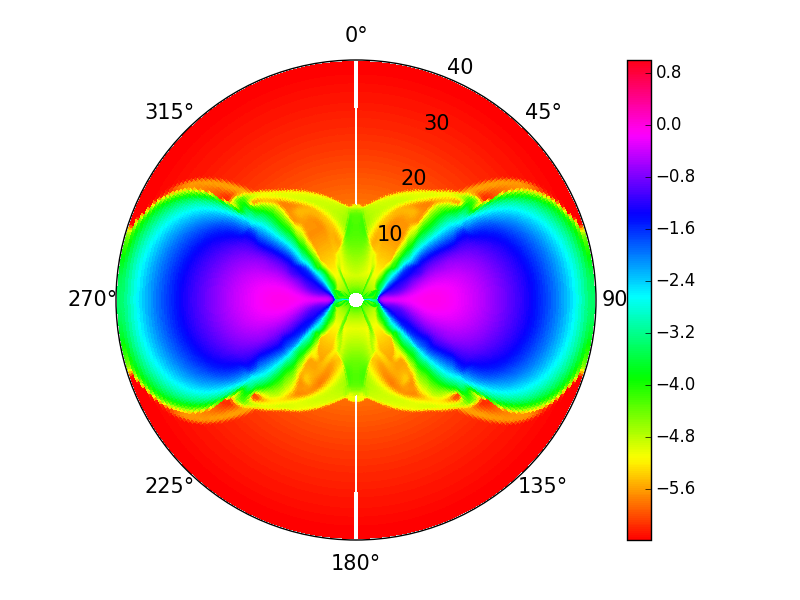}
\includegraphics[width=3.5in]{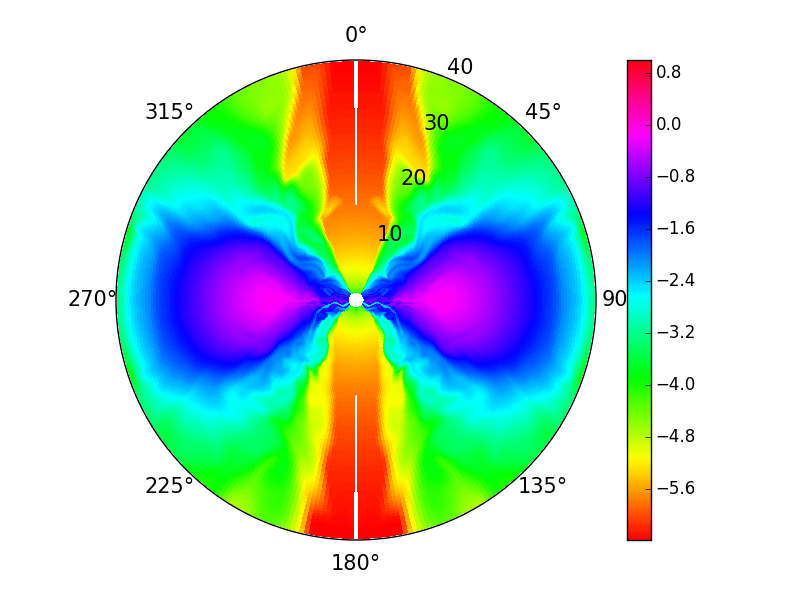}
\includegraphics[width=3.5in]{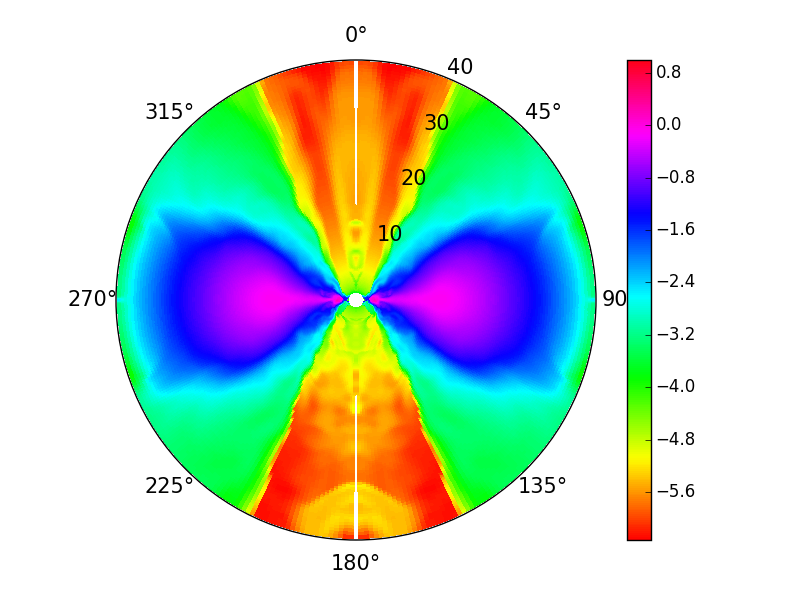}
\caption{The figure presents $lg(\rho)$ of simulations {\em torT0} (left column) and {\em torT2} (right column) at time $t=0$(top), $350t_{\rm g}$(middle), 
$700t_{\rm g}$(bottom). In simulation {\em torT0}, MRI made the torus unstable and later it became turbulent with an accretion flow
\citet{2003ApJ...589..444G}. On the other hand, the torus structure of simulation {\em torT2} evolves in a much less turbulent 
way where MRI is largely damped by the magnetic diffusivity.}
\label{torus_4}
\end{figure*}

Having verified the validity of \HAR\ in the ideal GR-MHD regime, we compare the results from simulation {\em torT0} 
and {\em torT2} to see how magnetic diffusivity influences the torus evolution. 
Since simulation {\em torT2} lasted longer than simulation {\em torT1}, we plot the accretion rate from $t=0$ to 
$t=700t_{\rm g}$ and compare it to simulation {\em torT0}. 
Both plots are shown in Figure \ref{torus_3}. 
In simulation {\em torT0} (left plot), the equilibrium of the torus breaks slowly and at about $t=220t_{\rm g}$ the 
accretion started. 
The perturbation of the inflow (choked accretion) tended to be steady after $t=300t_{\rm g}$. 
The average value of the mass accretion rate in simulation {\em torT0} from $t=500t_{\rm g}$ to $t=700t_{\rm g}$ is 
about $-0.31$. 
However, in simulation {\em torT2} with $\eta=10^{-3}$ (right plot), there was no sign of massive accretion before $t\sim 550t_{\rm g}$. 
The presence of magnetic diffusivity suppresses MRI and thus the the angular momentum transport by allowing for
relative motion between matter and magnetic field (see Section \ref{torus_MRI_subsect} for a detailed discussion). 
This lack of coupling prevents the decay of the torus equilibrium state and hence only allows for inefficient 
accretion. 
This can be seen even more clearly in Figure \ref{torus_4}, where we plot the log of the density from simulations
{\em torT0} and {\em torT2} at different time. 
The evolution of the torus in simulation {\em torT2} is much smoother (less turbulent) then in simulation {\em torT1}.

%-------------------------------------------------------------------------------------------
\subsection{MRI evolution in a resistive GR-MHD torus}
\label{torus_MRI_subsect}
We have shown that the evolution of the MRI can be suppressed by magnetic diffusivity, and therefore influences the mass accretion 
rate from the torus to the black hole. 
In order to quantify the gradual influence of diffusivity, we have performed simulations considering various strength of
diffusivity $\eta$ (denoted by {\em mriT}-simulations). 
Here we present the results for $\eta$ ranging from $10^{-12}$ to $10^{-3}$. 

The diffusion rate will be of order $k^2\eta$ \citep{2000ApJ...530..464F}, where $k$ stands for wave number. 
According to \citet{1991ApJ...376..214B}, the MRI may grow only in a certain range of wave numbers $k \in [0,k_{\rm max}]$, in the linear MRI regime.
Furthermore, there exists a wave number $k_{\rm MRI}$ for which the MRI growth rate reaches a maximum (see \citealt{1992ApJ...400..595H} for the case of a Keplerian disk). 
A certain number of MRI modes can therefore be damped out when $k_{\rm MRI}^{2}\eta$ is comparable to the maximum growth rate of MRI. 
Moreover, for a large enough $\eta$, it is even possible to damp out most of the MRI modes in the linear evolution of MRI.

In the following, we apply the time evolution of the mass accretion rate as indicator of the MRI growth in the torus. 
The mass accretion rate is attributed to the turbulent angular momentum exchange, triggered by the MRI. 
Thus, when magnetic diffusivity damps the growth of MRI, the point in time when massive accretion will set in, is delayed.
%Typically, the accretion rate is negligible for the initial few rotations until the MRI has grown substantially and angular momentum exchange may set in after several hundreds of time steps 
Also, the MRI in \citet{1991ApJ...376..214B} went into non-linear regime after about 2 rotations. In our simulations, the rotation period of the torus at pressure maximum is $267t_{\rm g}$ and about $98t_{\rm g}$ at the inner edge of torus. Thus, we assume that the growing of MRI becomes non-linear after $t=530t_{\rm g}$.

\begin{figure*}
\centering
\includegraphics[width=7.2in,height=2.4in]{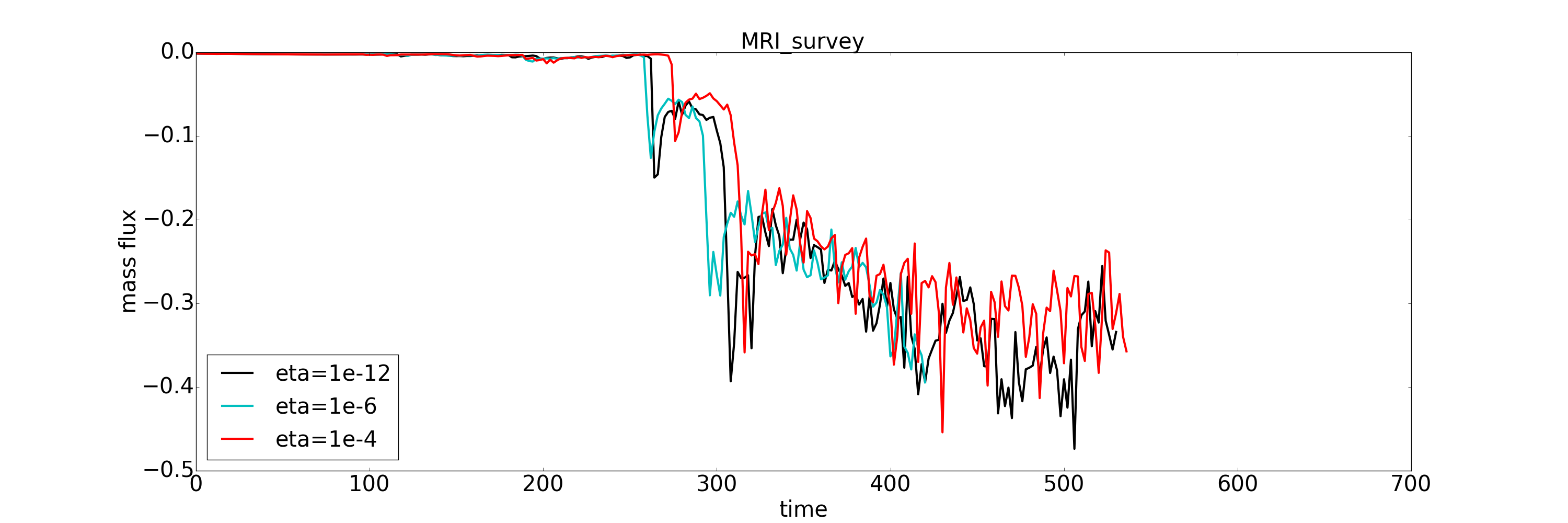}
\includegraphics[width=7.2in,height=2.4in]{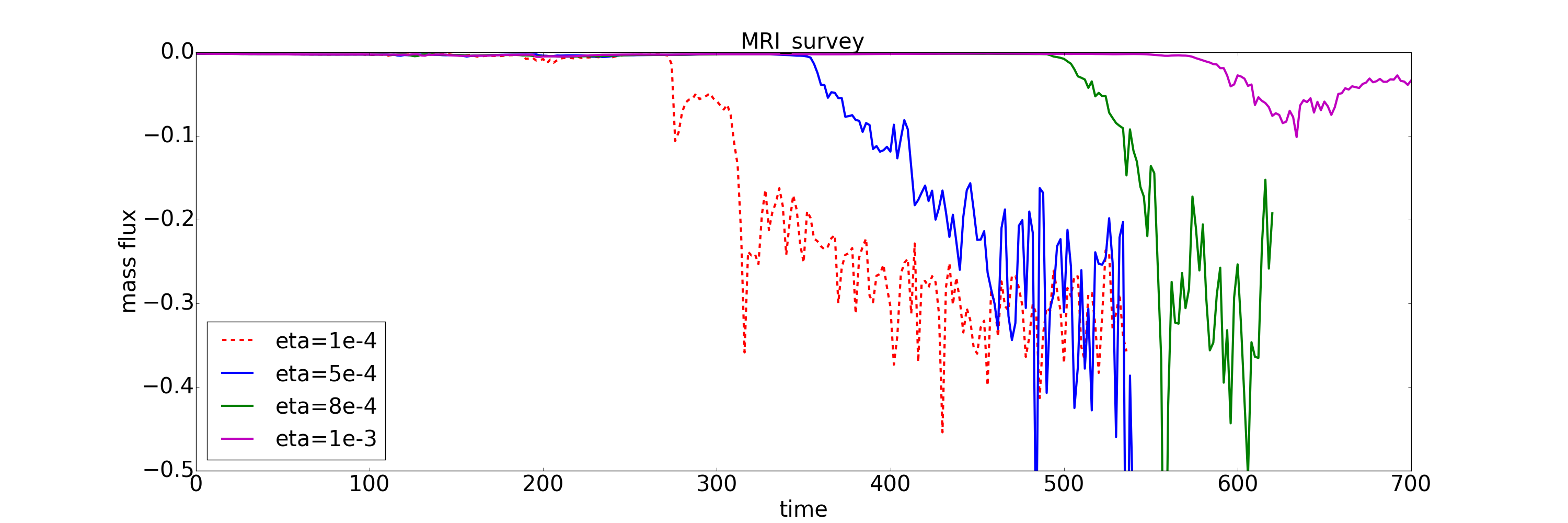}
\caption{Mass flux as the indicator of MRI growth in simulation {\em mriT1}-{\em mriT6}. 
Two plots are shown in order to avoid confusion between too many curves.
In the upper plot are the mass accretion rates for $\eta=10^{-12}$ (black), $\eta=10^{-6}$ (cyan) 
and $\eta=10^{-4}$ (red) while in the lower plot there are accretion rates for $\eta=5 \times 10^{-4}$ (cyan), 
$\eta=8 \times 10^{-4}$ (green) and $\eta=10^{-3}$ (magenta). The $\eta=10^{-4}$ (red dashed line) curve is 
plotted in the lower plot as a reference. 
Apparently, the time when substantial accretion initiates is delayed with increasing $\eta$.
%
%The critical value here is $\eta \approx 10^{-4}$. 
The delay in accretion can be explained by the magnetic diffusivity suppressing the MRI in the torus. 
For $\eta \leqslant 8 \times 10^{-4}$, only that part of the evolution is shown 
before these simulations experience numerical instabilities - similar to simulations {\em torT1} and {\em torT2}. 
}
\label{MRI_eta_plot}
\end{figure*}

The results of our simulations treating a magnetically diffusive torus are shown in Figure \ref{MRI_eta_plot}. 
We see that for this setup the accretion rate for the magnetic diffusivity $\eta = 10^{-6}$ does not distinguish from that
for $\eta = 10^{-12}$. 
In both cases, massive accretion takes place at about the same time compared the ideal GRMHD simulation {\it torT1}. 
Thus a small $\eta \leqslant 10^{-6}$ does not affect the growth of the MRI significantly.
Note that this is the range of $\eta$, for which simulations are probably dominated by numerical diffusivity 
(see Appendix \ref{num_diff_chapter} for a discussion on numerical diffusivity). 

On the other hand, for the simulations with $\eta \geqslant 10^{-4}$, the onset of massive accretion is
delayed. 
%The turn-over point lies some where between $10^{-4}$ and $10^{-3}$.
%After having run simulations for a range of magnetic diffusivity, 
We find indication for a critical value of $ 8\times10^{-4} \simeq \eta_{\rm crit} \simeq 10^{-3}$ 
for the magnetic diffusivity concerning MRI growth or mass accretion.
This value, of course, holds for the choice of our simulation setup, grid resolution, magnetic field strength, 
respectively. 
While for $\eta \gtrsim 5 \times 10^{-4}$ there is an obvious delay for massive accretion,
for $\eta > 10^{-3}$, the MRI seems to be completely suppressed during the linear regime
and for this parameter regime.
%(for our simulation setup, for grid resolution and magnetic field strength) and
%for the magnetic diffusivity concerning MRI growth or mass accretion, respectively.
This result is consistent with \citet{2010A&A...516A..51L}, who demonstrated that the growth rate of the MRI substantially decreases with $1/\eta$ as soon as critical diffusivity is exceeded.

\section{Winds launched from magnetically diffusive \& thin accretion disks}
In this section we present preliminary results of disk winds launched from magnetically diffusive accretion 
disks around black holes.
The launching of outflows from disks essentially lives from the existence of a large scale poloidal magnetic 
field threading the disk and the existence of a magnetic diffusivity in the plasma that allows both the 
accretion of matter across the field and the re-distribution of mass flux from accretion into ejection.
Numerical investigations of the launching of disk winds, thus the simulations of the accretion-ejection
transition have mostly been done for non-relativistic systems
(see e.g. \citealt{2002ApJ...581..988C, 2007A&A...469..811Z, 2009MNRAS.400..820T,2012ApJ...757...65S, 
2014ApJ...793...31S, 2015ApJ...814..113S}).
Here we extend this treatment to disks in GR-MHD for the first time.

\subsection{Simulation setup}
The numerical setup of these simulations is similar to the previous models discussed above - the same grid setup
and boundary conditions.
So far we have investigated only Schwarzschild black holes, the event horizon of which is at $r_{\rm H}=2r_{\rm g}$.
Obviously, we apply a different initial condition, that is a thin disk in Keplerian rotation threaded by a 
large scale magnetic field. 
These are the standard initial conditions for the non-relativistic launching simulations cited above.

For the disk initial velocity profile we assume a pure Keplerian rotation applying a 
Pacynzki-Wiita approximation for the disk \citep{1980A&A....88...23P} 
\begin{equation}
\Omega = r^{-3/2} \left(\frac{r}{r - R_{\rm pw}}\right),
\end{equation}
with a smoothing length scale $R_{\rm pw} = 2.0$ for convenience.
The Pacynzki-Wiita rotation profile has been chosen mainly for simplicity.
The system will anyway evolve into a new dynamical equilibrium thus a new disk gas
density and pressure distribution.
Another option would have been a true GR solution for the disk rotation curve.
However, also in that case we would be left with the question of the initial disk density 
and pressure distribution that are a priori unknown.
For our simulations we choose an inner disk radius of $r_{\rm in} = 3 r_{\rm H} = 6 r_{\rm g}$ initially.

For the disk density and pressure distribution and the initial poloidal magnetic field
we apply the typical choice for the non-relativistic simulations, thus
\begin{equation}
\rho (r,\theta) = 
\frac{R_{\rm pw}^3}{\left(R_{\rm pw}^2 + r^2 \right)^{3/2} } 
\left(1 - (\gamma_{\rm G} - 1)\frac{r^2\cos^2\theta}{2 \epsilon_{\rm D}^2 r^2} \right)^{1 / (\gamma_{\rm G} -1)}
\label{eq:rho_thindisk}
\end{equation}
\citep{2002ApJ...581..988C}, 
with the classical disk aspect ratio $\epsilon_{\rm D} \equiv H/r$ and the local 
disk height $H(r)$. The gas pressure follows $p = K \rho^{\gamma_{\rm G}}$.
Equation (\ref{eq:rho_thindisk}) describes a vertical density and pressure profile that steeply
decreases with distance from the disk midplane. 
In difference to the simulations in the Newtonian limit cited above 
(see e.g. \citealt{2002ApJ...581..988C,2007A&A...469..811Z}, 
where a non-relativistic gas with $\gamma_{\rm G} = 4/3$ for the gas polytropic index, here
we consider a relativistically hot gas with $\gamma_{\rm G} = 4/3$.

For the initial poloidal magnetic field we apply the vector potential
\begin{equation}
A(r,\theta) = \frac{5}{2} B_{\rm p,0}  
              \left(r \sin\theta\right)^{3/4}  
              \frac{m^{5/4}}{\left( m^2 + \tan^{-2}\theta \right)^{5/8}},
\label{eq:Binit}
\end{equation}
 \citet{2007A&A...469..811Z, 2012ApJ...757...65S}.
The parameter $B_{\rm p,0}$ determines the strength of the initial magnetic field and is determined by the 
choice of the plasma beta, while the parameter $m$ defines the opening angle of the magnetic field lines 
(typically $m=0.4$).

For the magnetic diffusivity we follow the non-relativistic simulations such that we imply 
a turbulent magnetic diffusivity within the disk and the disk corona.
The vertical profile of the magnetic diffusivity decreases exponentially with distance from the disk 
midplane,
\begin{equation}
    \eta(r,\theta) = \eta_0 r^{-1/2}  \exp\left[- 2 \left(\frac{\alpha}{\alpha_{\eta}}\right)^2\right],
\end{equation}
where $\alpha \equiv \pi/2 -\theta$ is the angle towards the disk {\em midplane}, 
and $\alpha_{\eta} \equiv \arctan(H_{\eta} \epsilon_{\rm D})$
is the angle defining the scale height of the diffusivity profile via $H_{\eta}$.
For example, for $H_{\eta} = 3$, the scale height of diffusivity is three times larger than the disk pressure scale height
(see e.g. discussion in \citealt{2012ApJ...757...65S}).

\begin{figure*}
\centering
\includegraphics[width=1.7in]{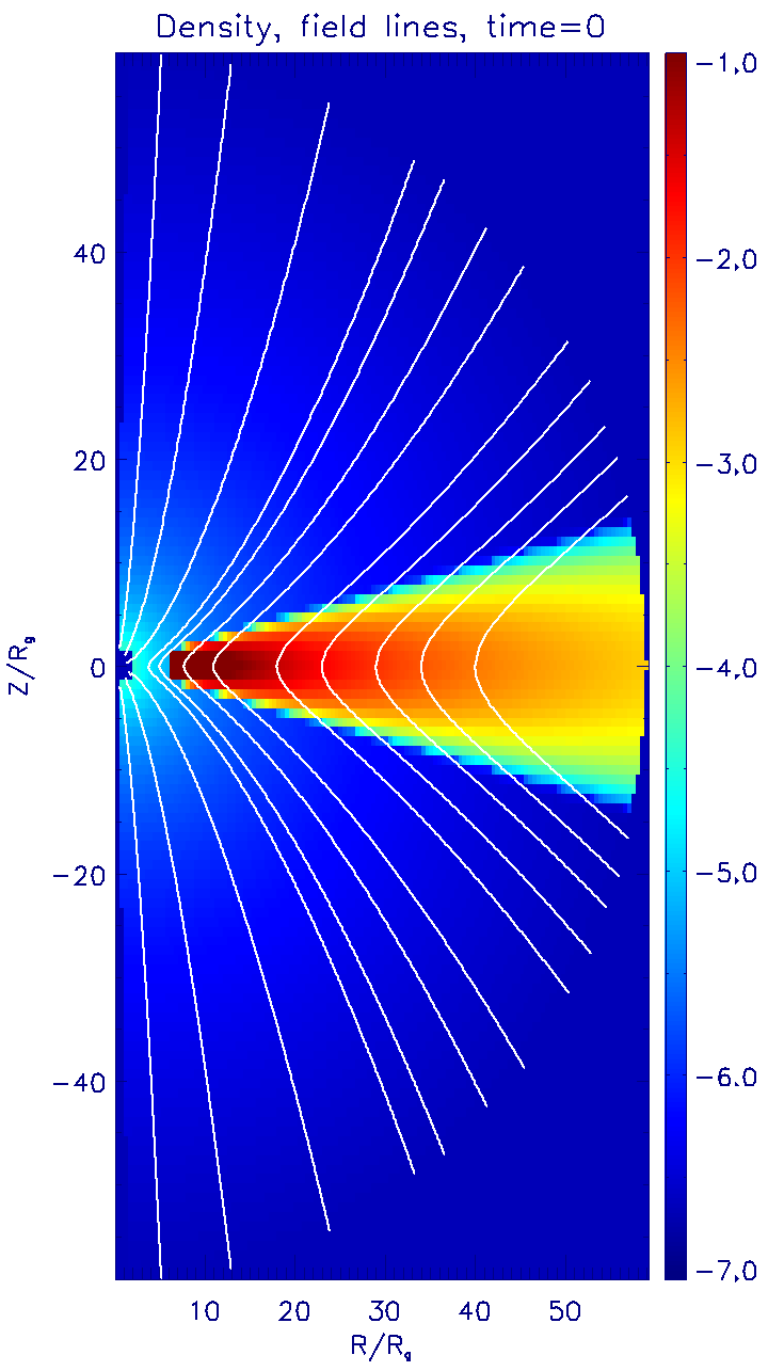}
\includegraphics[width=1.7in]{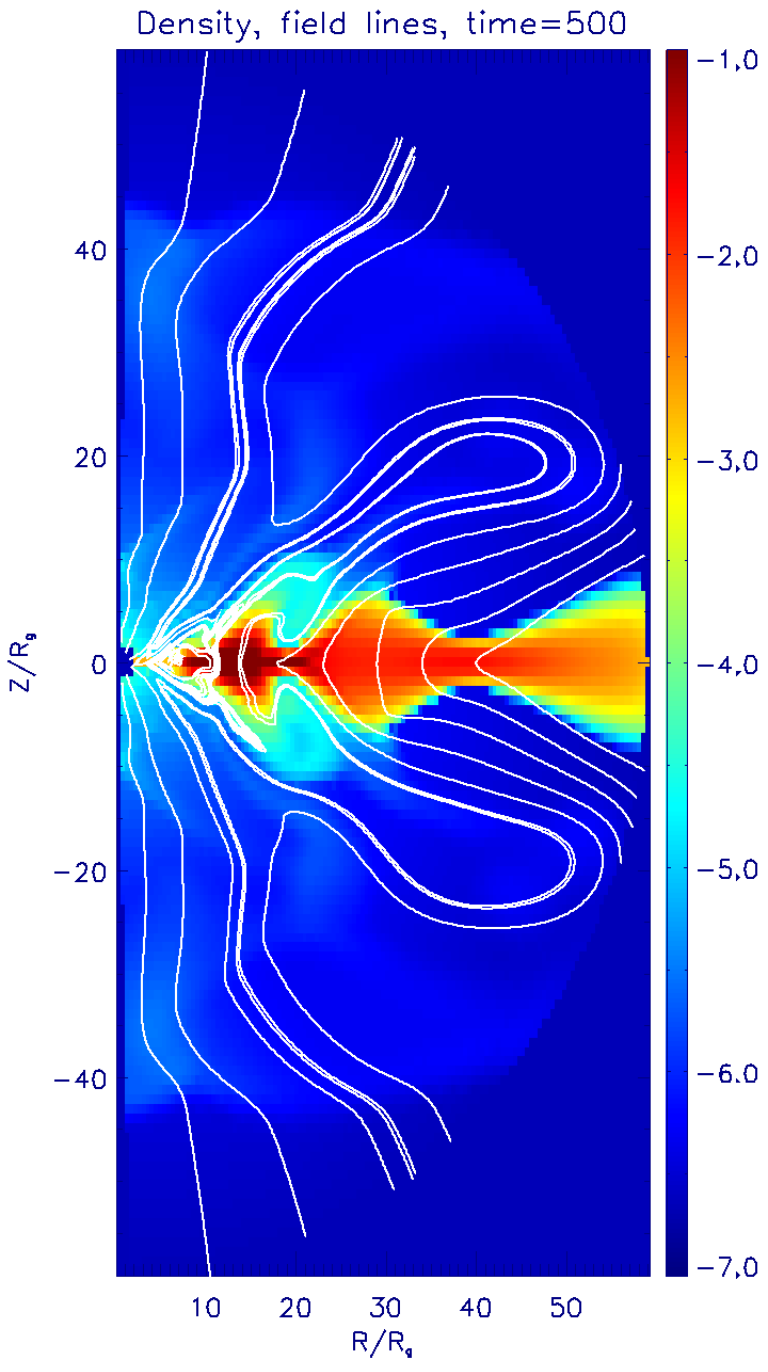}
\includegraphics[width=1.7in]{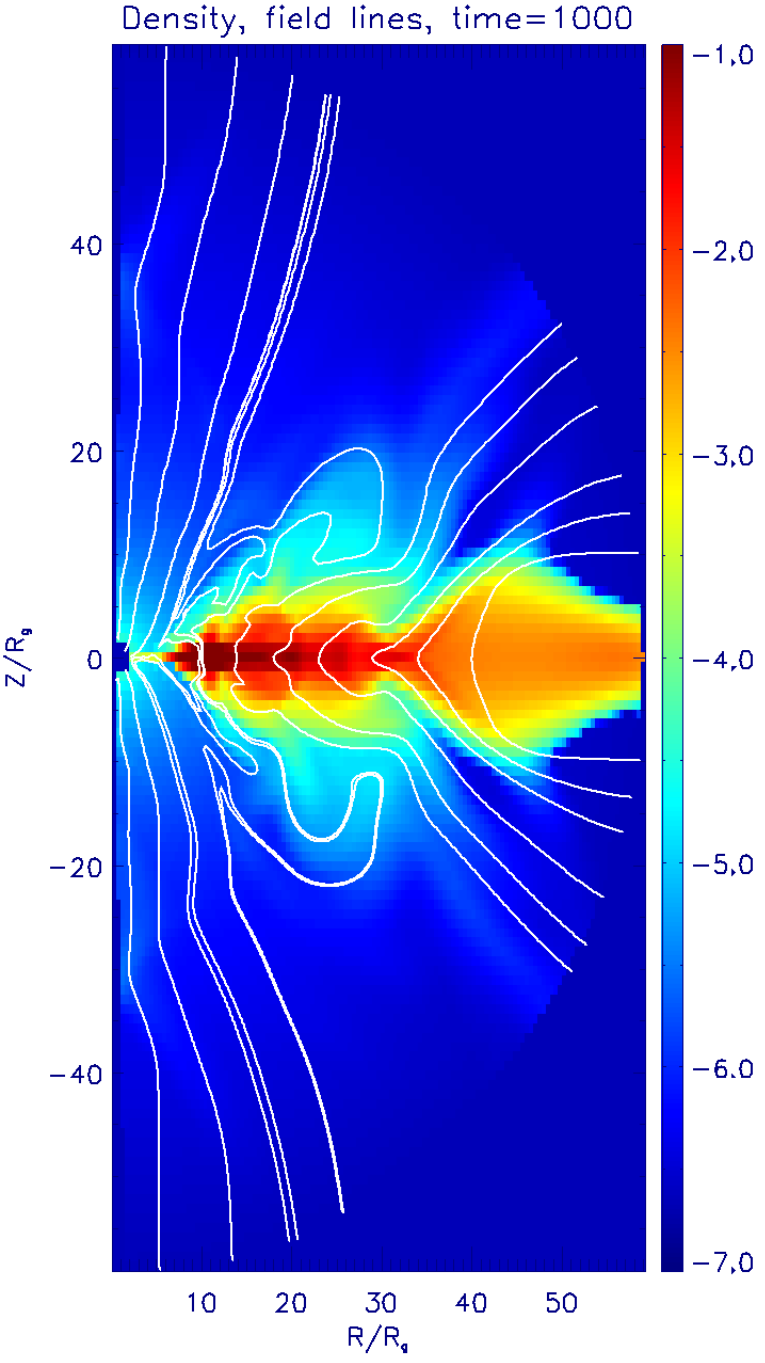}
\includegraphics[width=1.7in]{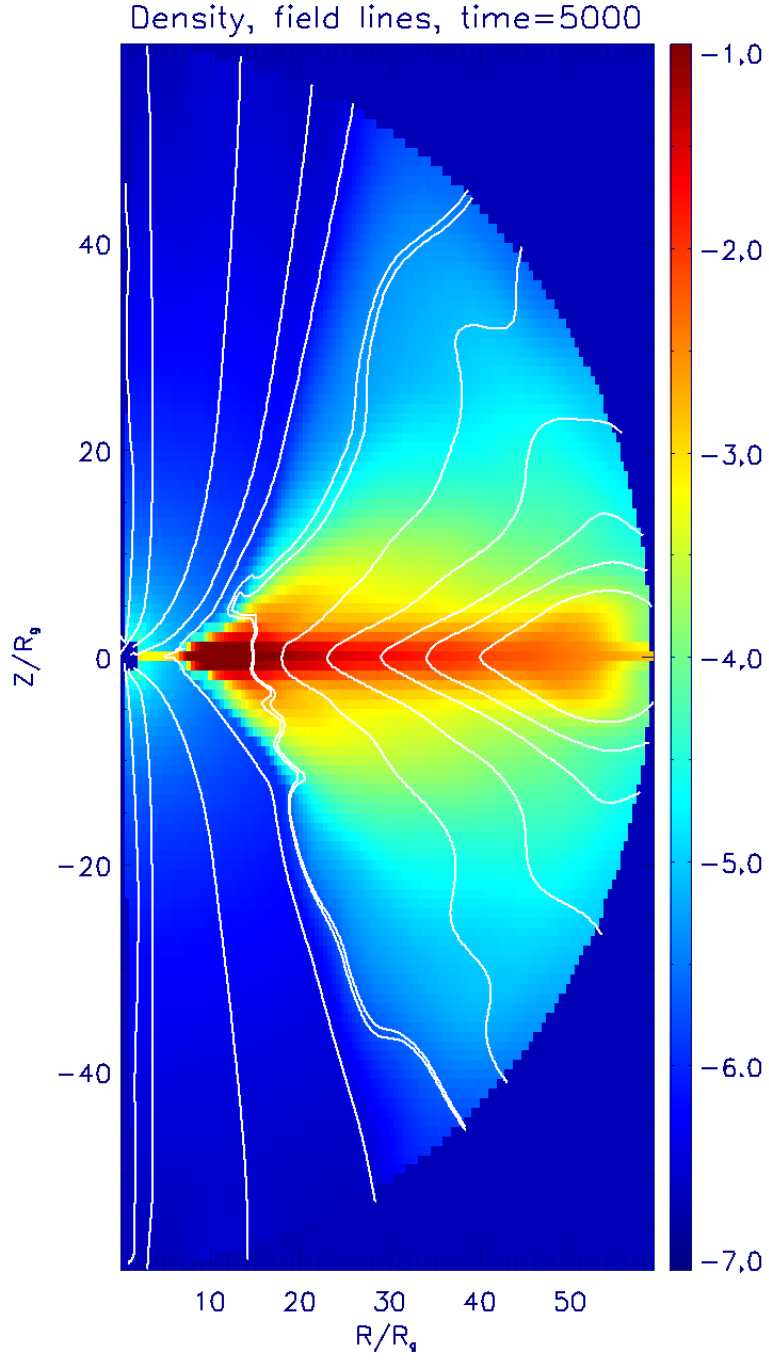}
\caption{Winds launched from resistive MHD disks. 
Shown is the density distribution (color coding) and the magnetic field lines (white lines) for time steps
$t=0, 500, 1000, 5000$.
}
\label{disk_d_phys}
\end{figure*}

\begin{figure*}
\centering
\includegraphics[width=1.5in]{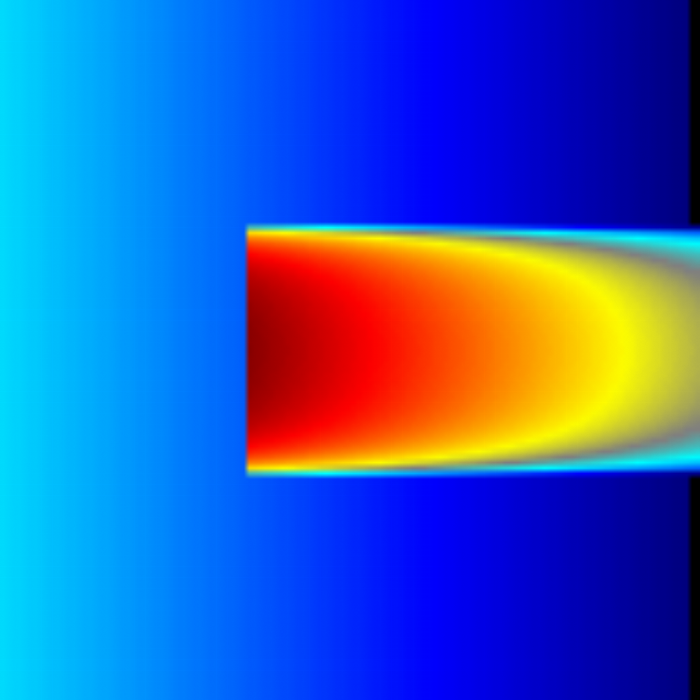}
\includegraphics[width=1.5in]{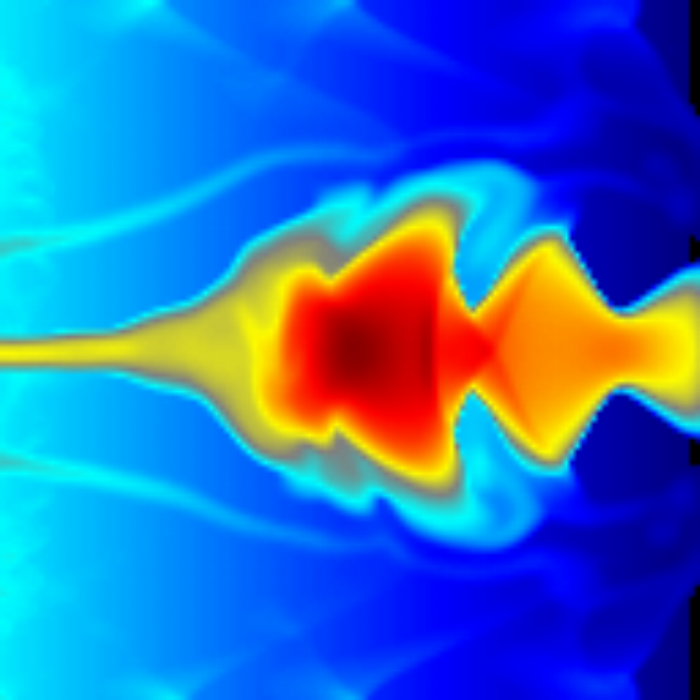}
\includegraphics[width=1.5in]{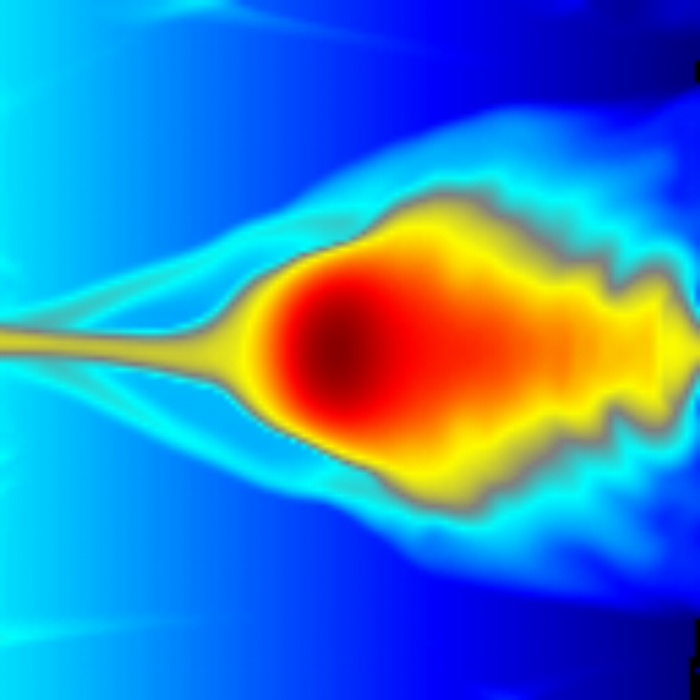}
\includegraphics[width=1.5in]{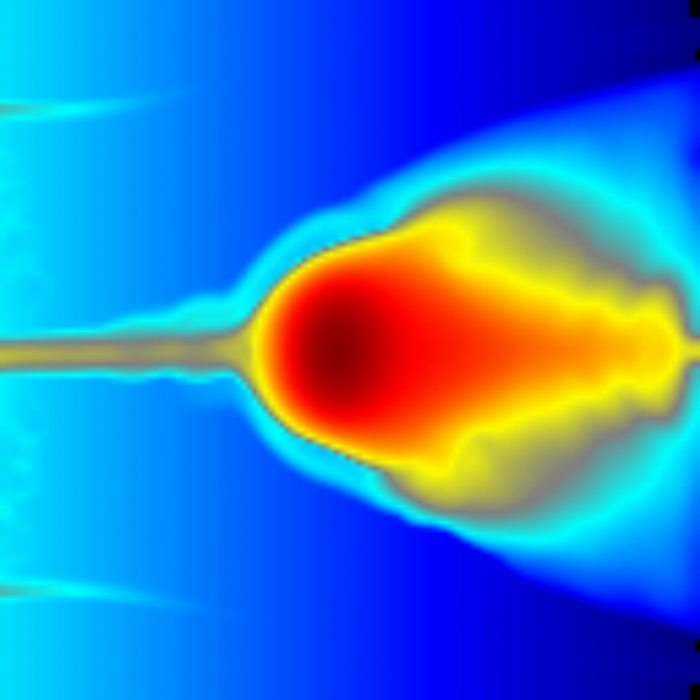}
\caption{Winds launched from resistive MHD disks. 
Shown is the density distribution (color coding) on grid coordinates, demonstrating the
small vertical extension of the final accretion stream.
We further see that this stream is resolved vertically by about 10 grid cells, while the inner
disk is resolved by 20-50 grid cells.
}
\label{disk_d_grid}
\end{figure*}

\begin{figure*}
\centering
\includegraphics[width=3.5in]{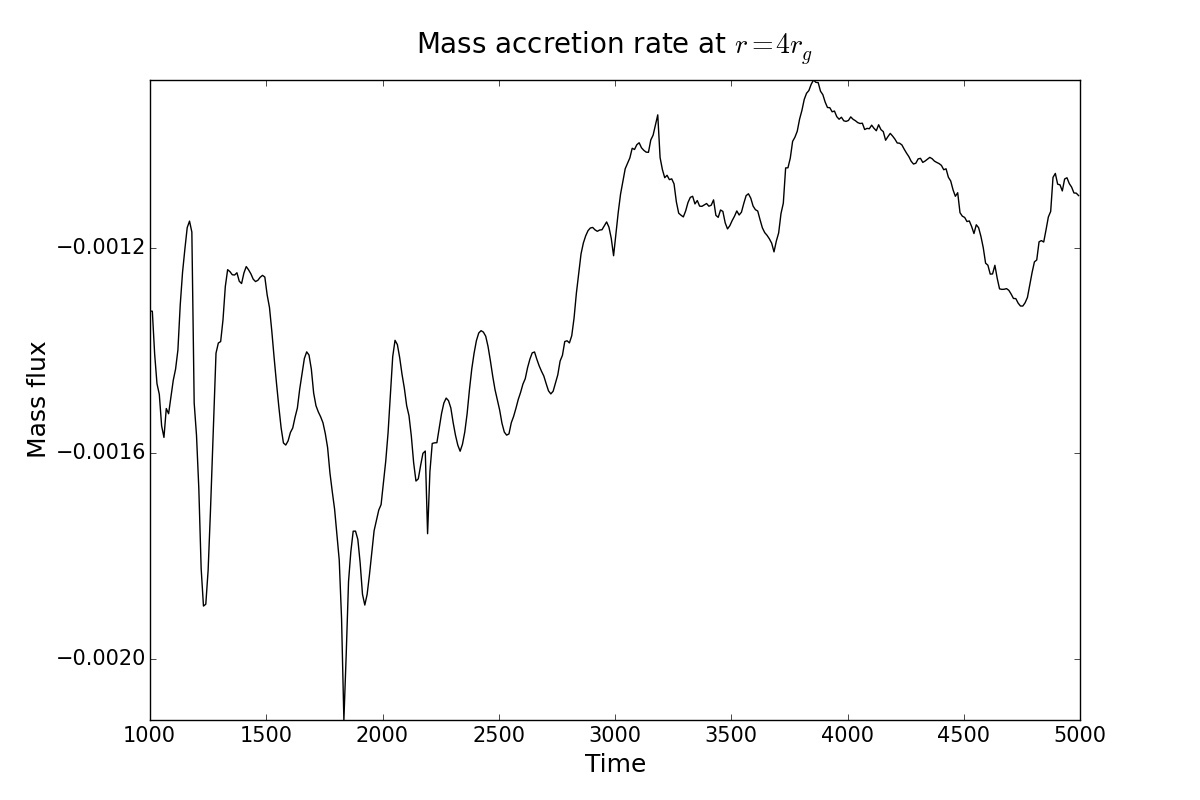}
\includegraphics[width=3.5in]{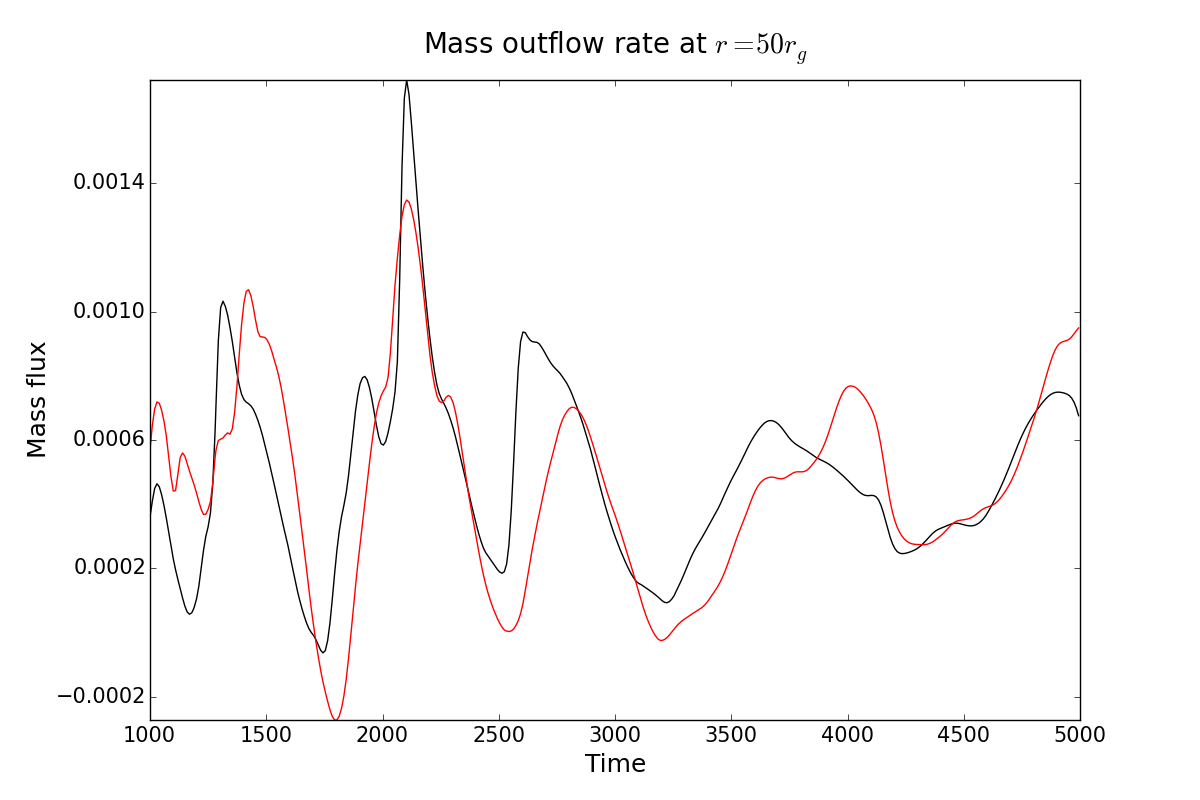}
\caption{Winds launched from resistive MHD disks. 
Shown are the mass fluxes for accretion and ejection.
The mass accretion (left) is integrated at $r=4$ and for $70^o < \theta < 110^o$.
The outflow rates (right) are integrated along a circle with $r=50$ and for
$30^o < \theta < 70^o$ (black line, upper hemisphere) and
$110^o < \theta < 150^o$ (red line,  lower hemisphere), respectively.
Note the unsteady character of accretion and ejection and the symmetry
in the outflow rates of the upper and lower hemispheres.
In this simulation, the total outflow is comparable to the accretion rate.
% This is simulation: /HARM-QQ-NEW/Standard/Run09_Christos_paras/Sim03
}
\label{disk_mfluxes}
\end{figure*}

\begin{figure*}
\centering
\includegraphics[height=4.in]{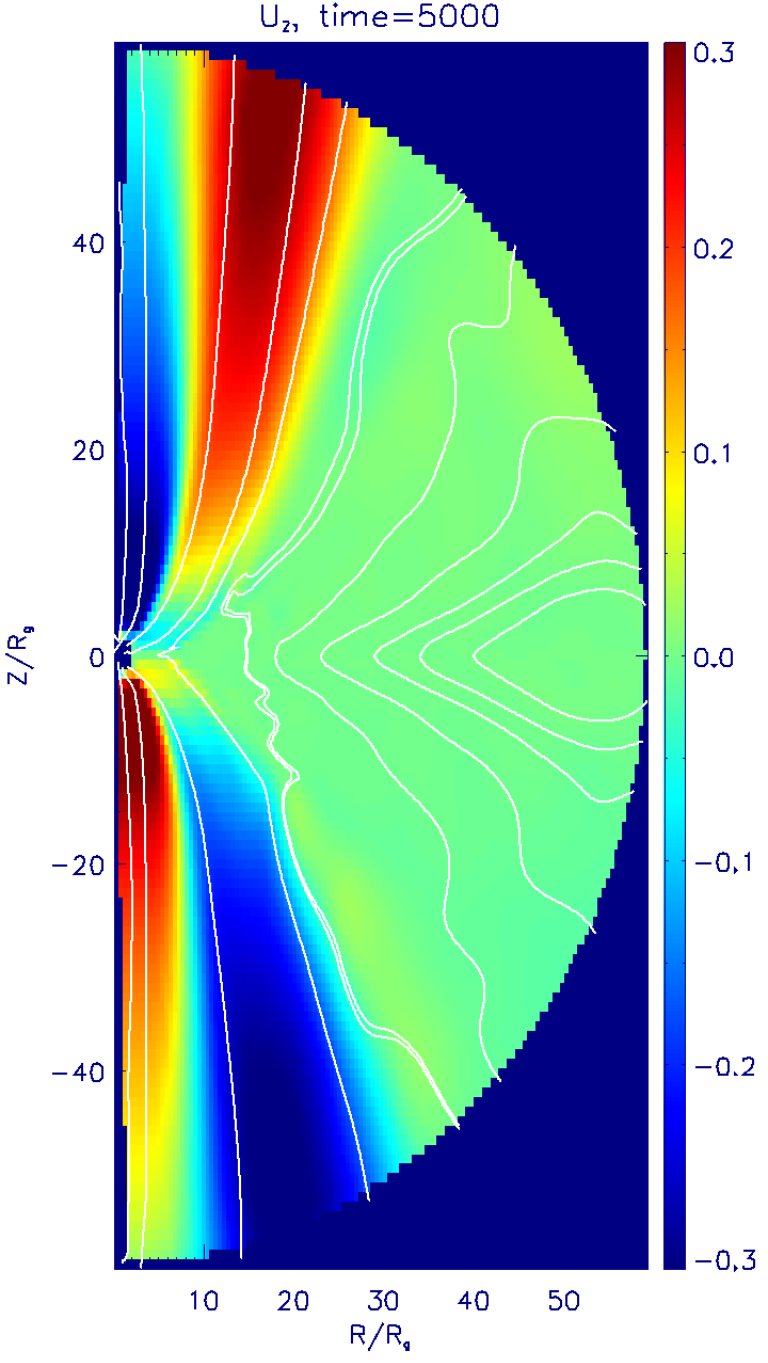}
\includegraphics[height=4.in]{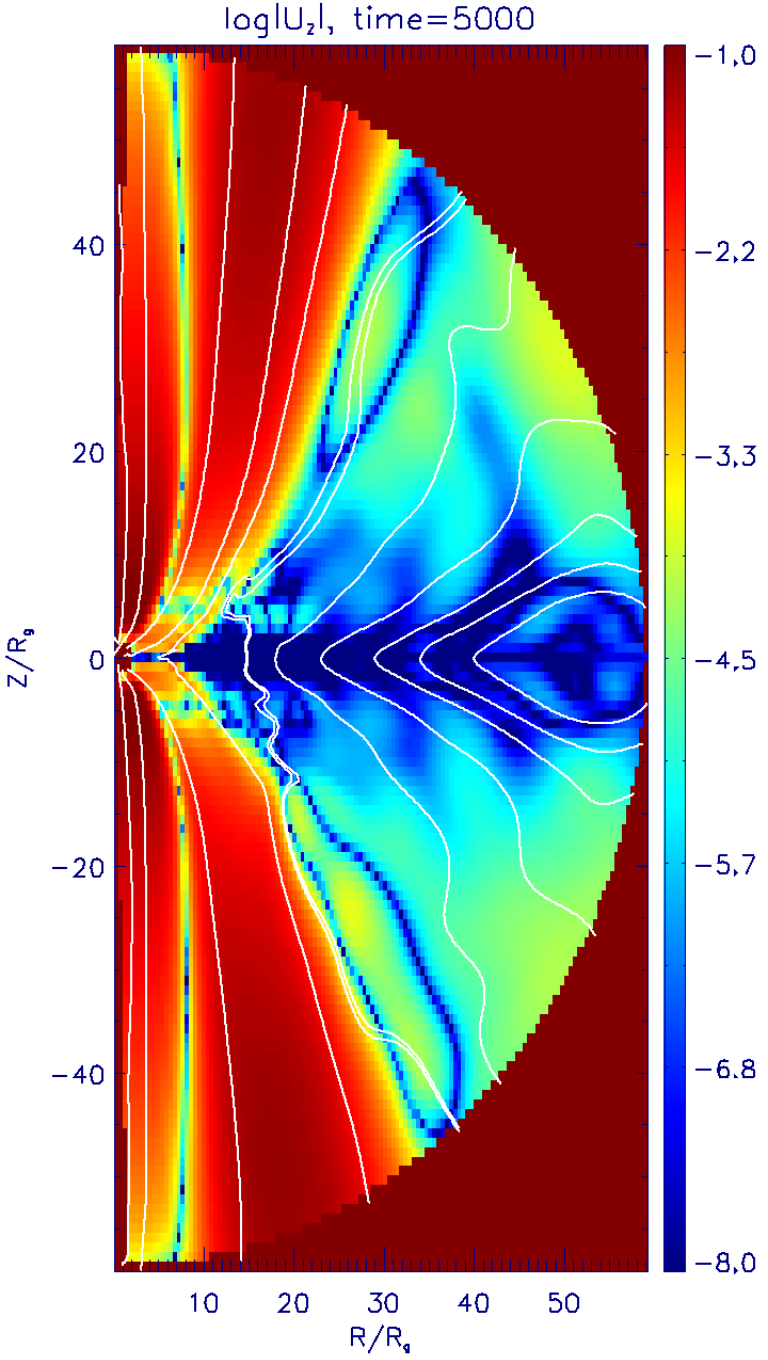}
\includegraphics[height=4.in]{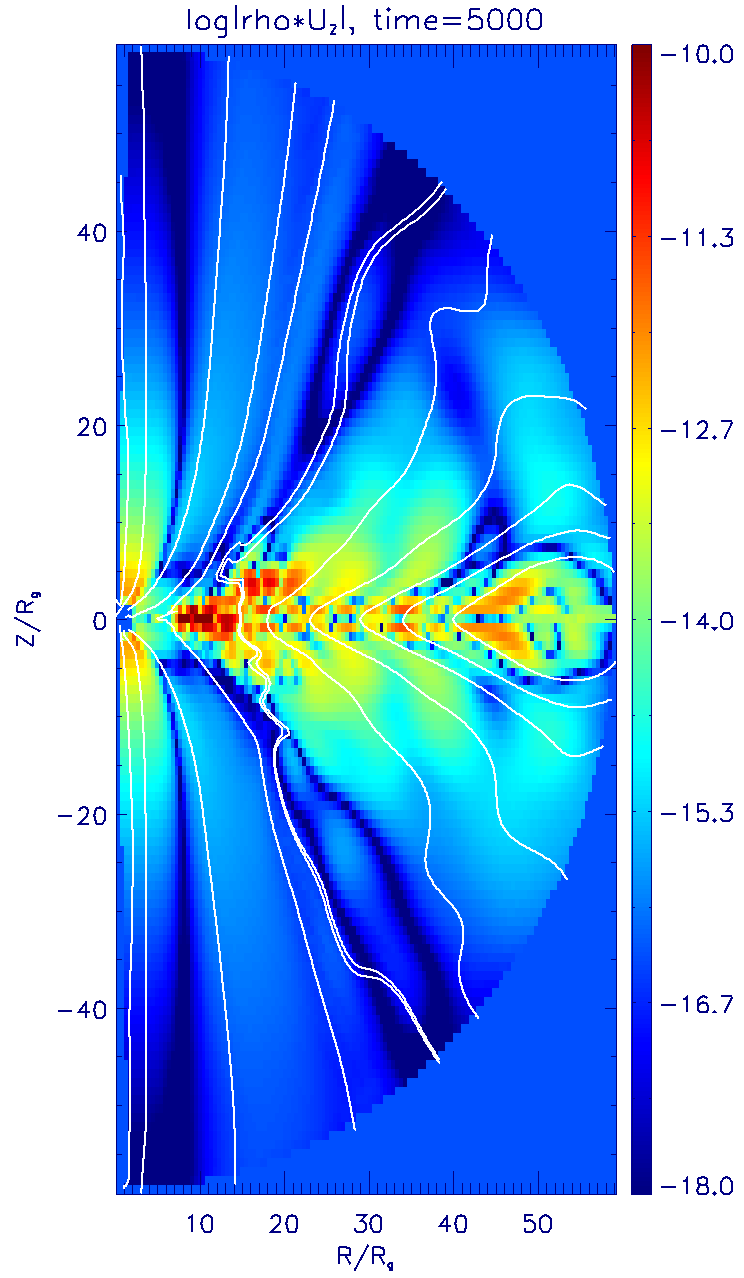}
\caption{Winds launched from resistive MHD disks. Shown are the vertical velocity and mass flux distribution with magnetic field lines at $t=5000$. Linear scaling for the vertical velocity $u_z$ (left), log scaling for the vertical velocity $\log|u_z|$ (middle), and local mass flux mass accretion $\log|\rho u_z|$ (right).
}
\label{fig_velo}
\end{figure*}

\subsection{Simulation results}
We now discuss an example simulation of a thin disk simulation. 
The parameters of this simulation are the following: 
$a=0$, $K=0.01$, $\beta = 100$, 
$m=0.4$, $R_{\rm in} = 0.85 r_{\rm H} = 1.7 r_{\rm g}$,
$r_{\rm out} = 60 r_{\rm g}$, 
$\epsilon_{\rm D} = 0.1$, 
$\eta_0 = 10^{-3}$, $\eta_{\rm H} = 3$.
The simulation run for 5000 time units, corresponding to about 80 rotations of the inner disk (at $r=4 r_{\rm g}$). 
Figure \ref{disk_d_phys} shows the density structure and the magnetic field lines of the simulation for times
$t=0, 500,1000,5000 t_{\rm g}$ in Boyer-Lindquist coordinates.

We see that the initial disk structure becomes quite heavily disturbed. 
Accretion shocks occur, however a new dynamical equilibrium is established for the inner disk after about $t=1000$.
For the outer disk it takes longer, just because the natural time scale, the Keplerian time scale is longer.
At the end of the simulation (that we stopped at $t=5000 t_{\rm g}$) we see a disk wind emerging from all over the disk surface.
This disk wind has a rather smooth density structure.
The disk wind encloses an inner, axial cone of (half) opening angle of about $30^o$.
No material is ejected from this area, as we consider a Schwarzschild black hole that does not provide driving of
an outflow.

Towards the end of our simulation we see that the outer disk structure begins to disappear.
This is partly due to the outflow boundary condition set at this position, but also due to the
mass loss by accretion and ejection (see below for numerical values).
A larger mass reservoir would be needed for longer lasting simulations.

%{\rot Speed of outflow? No collimation yet.}

In Figure \ref{disk_d_grid} we show the density evolution on the grid coordinates.
Note that in this representation the left boundary represents the inner boundary inside the horizon,
the upper and lower boundaries represent the rotational axis, and the right boundary represents the
outer circular outflow boundary reaching from the upper to the lower symmetry axis, thus from
$\theta = 0$ to $\theta = \pi$.
The left figure clearly shows the inner disk radius of the initial condition, chosen to be at 
$3 r_{\rm H} = 6 r_{\rm g}$,
thus at the marginally stable orbit.
During the simulation, the disk inner radius moves to a slightly larger radius of $r\simeq 4 r_{\rm H}$ 
(denoted by the red density contours), and stays constant for the rest of the simulation time.
Between the accretion disk structure and the horizon a thin accretion stream evolves extending only few degrees in vertical direction.
The (inner) disk is resolved with 25 grid cells per disk scale height initially. 
While the disk resolution stays about the same with time, the fast accretion stream that develops  
between inner disk and black hole is resolved with about 9 grid cells vertically.
%The {"}opening angle{"} is about 10 grid cells in $\theta$ corresponding to an angle of XXX degrees (note the "hslope").
%mark

We now have a closer look to the velocity distribution (see Figure \ref{fig_velo}).
Our simulation shows a two-component outflow structure.
A high velocity outflow is launched from the disk very close to the black hole ($r<10r_{\rm g}$).
Yet the velocities are mildly relativistic due to the choice of the magnetic field strength
that is not very strong with $\beta = 100$.
This high speed outflow is fed with disk material due to magnetic resistivity
and is rotating rapidly (not shown).
Outside the high speed flow, a rather low-speed disk wind emerges, that carries, 
however, somewhat more mass.
The low velocity mass flow is highly structured, in difference to the high-speed 
outflow.
The low velocity outflow is aligned with the disk magnetic field, however, it seems
to be driven by gas pressure gradient, and not by magneto-centrifugal effects. 
This is again a result of the low magnetic field strength initially assumed.
In a forthcoming paper we will investigate in detail the physics of these disk 
winds.

In Figure \ref{disk_mfluxes} we show the time evolution of the mass fluxes for different locations.
The left figure shows the mass accretion rate integrated at $r=4 r_{\rm g}$ (thus in the thin accretion stream),
for $70^o < \theta < 110^o$.
The accretion rate is quite variable, but fluctuating around a quite well defined average value.
The accretion rate decreases substantially after $t=3000 t_{\rm g}$. 
We believe that this happens because (i) the disk has lost mass substantially and changes its characteristics,
and because (ii) that disk wind is now well established and further contributes to the mass loss.
The average accretion rates we measure are $\dot{M} \simeq 0.0015$ for $t<3000 t_{\rm g}$ and $\dot{M} \simeq 0.0010$ 
for $t>3000 t_{\rm g}$. 

%\begin{figure*}[ht]
%\centering
%\includegraphics[height=4.in]{sim3_uz_t5000.png}
%\includegraphics[height=4.in]{sim3_log-uz_t5000.png}
%\includegraphics[height=4.in]{sim3_log-rho-uz_t5000.png}
%\caption{Winds launched from resistive MHD disks. 
%Shown are the vertical velocity and mass flux distribution with magnetic field lines at $t=5000$.
%Linear scaling for the vertical velocity $u_z$ (left), 
%log scaling for the vertical velocity $\log|u_z|$ (middle),
%and local mass flux mass accretion $\log|\rho u_z|$ (right).
% in Boyer-Lindquist coordinates.
% This is simulation: /HARM-QQ-NEW/Standard/Run09_Christos_paras/Sim03
%}
%\label{fig_velo}
%\end{figure*}

For the outflow mass fluxes we have measured the rates in the upper and lower hemisphere separately.
Again we see a unsteady behavior, however, now with some periodic structure of several 100 time units.
The time variation of lower and upper hemisphere is clearly correlated and must thus result from the
physical evolution of the disk structure.
The outflow mass fluxes we have integrated along a sphere with radius $r=50 r_{\rm g}$ (thus close to the outer boundary)
and for
$30^o < \theta < 70^o$ (upper hemisphere) and
$110^o < \theta < 150^o$ (lower hemisphere), respectively.
The average mass fluxes between $t=1000$ and $t=5000 t_{\rm g}$ are quite similar for both hemispheres in spite of the
large intrinsic fluctuations.
For the upper hemisphere we measure an average outflow rate of
$\dot{M} \simeq 0.00049$, while for the lower hemisphere we find $\dot{M} \simeq 0.00046$.
We believe that the reason for this weak asymmetry lies in the fact that we consider
{\em resistive} MHD simulations.
Resistivity leads to reconnection events that bring some statistical effect in the long term evolution.  

When we compare the mass fluxes of accretion and ejection we find that both rates are of the 
same order, 150\% of the ejection rate is accreted towards the black hole.
% Note, however, that the accretion rate at larger radii is larger, for $r=20 r_{\rm g}$ we
% measure $\dot{M} \simeq 0.002$. % check number again XXXXXX
%
% The difference of the accretion rates for the two radii must be ejected in the outflow.

A more complete investigation would have to compare the accretion and accretion rates for different
radii and see how the ratio of mass fluxes will change along radius.
We defer such investigations to a follow-up paper devoted solely to thin disk in resistive GR-MHD.

%==================================================================================================
\section{Summary}
In this work, we have implemented resistivity, respectively magnetic diffusivity into the ideal GR-MHD code HARM 
\citep{2006ApJ...641..626N}. We denote the now code as \HAR.
Our paper illustrates the implementation and provides test simulations as well as preliminary
astrophysical results.

The implementation of resistivity applies the general definition of Faraday tensor - hence the 
general form of the stress energy tensor including the electric field - to the new code \HAR.
We follow the equations in \citet{2013MNRAS.428...71B} to calculate the electric field. 
Our inversion scheme that is based on the 2D inversion scheme in \citet{2006ApJ...641..626N} 
uses an extra loop to make the electric field variables converge.

We have verified our implementation of resistivity in \HAR\ by comparing the diffusion of an initial 
magnetic field distribution to the analytic time of the profile as given by the diffusion equation.
These simulations were performed in rectangular boxes of weakly magnetized gas, excluding any
dynamical effect by Lorentz forces.
Boxes at different distance from the black hole were investigated.
The magnetic diffusion evolving in \HAR\ are identical to the known analytic solution for
different magnetic diffusivities from $\eta=10^{-10}$ to $\eta=10^{-2}$.

We have further tested \HAR\ by a classical shock tube problem, finding very good agreement for magnetic 
diffusivities $\eta <0.1$.
For larger diffusivity, \HAR\ does not capture the shock front perfectly anymore, but such large
diffusivities are beyond the scope of of our aims of treating the disk accretion-ejection structure.

Having implemented physical magnetic diffusivity in the code, we are now able to measure the
numerical diffusivity.
That clearly depends of setup and resolution, but for a cell size of $\Delta x \simeq 0.01$ we find the
numerical diffusivity 2-3 orders of magnitude below the physical diffusivity applied in our accretion disk
setup.

We have then applied \HAR\ in a more astrophysical context. 
We have investigated (i) the development to the magneto-rotational instability (MRI)
in tori that are magnetically  diffusive, and (ii) the launching of disk winds from 
thin disks.
%In the simulations under astronomical enviroment. 
For the MRI simulations we applied an initial setup as in \citealt{2003ApJ...589..444G, 2006ApJ...641..626N}
that is an initially stable gas torus that carries a poloidal magnetic field that follows the density contours.

First, as a further verification of the new code \HAR, we have run a simulation with a very small magnetic  
diffusivity $\eta=10^{-12}$ (that is clearly below the numerical diffusivity of the code).
This simulation recovers the time evolution of the accretion rate that has been found 
previously by using the ideal GR-MHD code HARM.
In contrast, in the simulation with a high diffusivity $\eta=10^{-3}$, the mass accretion onto the black 
hole decreases significantly due to the suppression of MRI. 

In order to investigate further the influence of magnetic diffusivity on relativistic MRI tori, 
we have performed a parameter survey ranging from $\eta=10^{-12}$ to $\eta=10^{-3}$.
We find indication for a critical value for the magnetic diffusivity of $\eta \gtrsim 5\times10^{-4}$ 
(in this specific simulation model), above which the MRI is suppressed in the linear regime.

We finally presented preliminary results of MHD launching of disk winds from thin accretion disk
threaded by inclined open poloidal field lines.
Magnetic diffusivity allows to exchange mass between magnetic flux surface, and thus to load the open field
lines with material from the accretion disk.
The simulations did not yet reach a quasi-steady state of accretion-ejection that is know from
non-relativistic simulations.
However, the average mass accretion-to-ejection rate is similar, but somewhat larger compared
to non-relativistic studies and can reach values up to unity at some times.
We will present further investigations of the disk-driven winds in a follow-up paper.

%==========================================================================================
\acknowledgements

We thank Niccol\`o Bucciantini and Luca Del Zanna for insights concerning the implementation
of resistivity in relativistic MHD.
We thank Jonathan McKinney for a helpful comment on the 1D shock tube setup.
We acknowledge test simulations of the initial conditions for the thin approach disk by Christos 
Vourellis.

%==========================================================================================

\appendix

\section{ A. THE NUMERICAL PROCEDURE OF THE TIME EVOLUTION IN \HAR}
\label{num_pro_chapter}
\begin{figure*}[ht]
\centering
\includegraphics[height=5.in]{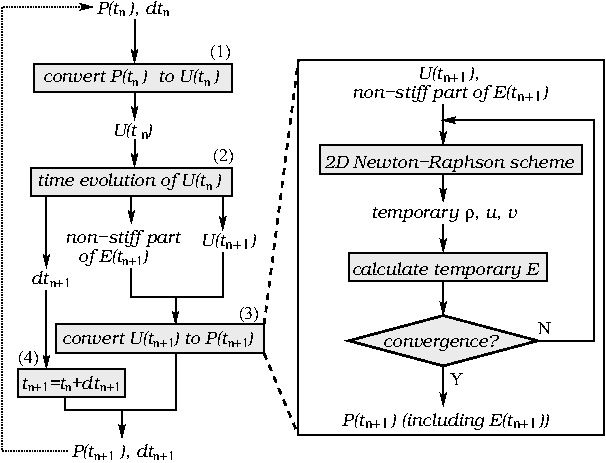}
\caption{The flow chart describing the numerical procedure of one time evolution in \HAR. Boxes with grey background denotes one or a series of routines that achieve the function written inside these boxes. The flow chart on the left hand side presents the procedure of one time evolution while the large box on the right hand side is a detailed flow description for the process inside the box (3) on the left hand side. See the text for explanations of each step in the chart.
}
\label{flow_chart}
\end{figure*}

The numerical procedure of one step in time evolution in \HAR\ follows from the derivations in Section \ref{theo_eqs_chapter} and \ref{inversion_scheme_chapter}. 
For further understanding we present these procedures in a flow chart (see  Figure \ref{flow_chart})
with the following explanations to each step shown in the chart.
\begin{enumerate}

 \item[(1)] We take the primitive variables $\pmb{P}(t_{n})$ from the previous step $t_{n}$ and convert them to the conserved variables $\pmb{U}(t_{n})$ (see Section \ref{inversion_scheme_chapter} for the definitions of $\pmb{P}$ and $\pmb{U}$).
 The $D$ can be calculated by $D \equiv \rho u^{t}$ as defined by Equation~\ref{conserved_variables}.
 The $T^{t}_{\,\,\,\mu}$ can be obtained with the help of Equation~\ref{eq_str_ene_tens_2}. 
 The magnetic and electric fields $\alpha ^\ast F^{it}, -\alpha F^{it}$ (hereafter $\pmb{\mathcal{B}}$, $\pmb{\mathcal{E}}$, see Section \ref{theo_eqs_chapter}) are already provided, since they are both conserved and primitive variables.
            
 \item[(2)] We evolve the conserved variables from $U(t_{n})$ to $U(t_{n+1})$ (except $\pmb{\mathcal{E}}$). 
 To do that, we first need to calculate the flux of $U(t_{n})$. 
 Knowing $\pmb{P}$, the flux of $D \equiv \rho u^{t}$ is $\rho u^{i}$ and the fluxes of $T^{t}_{\,\,\,\mu}$ are $T^{i}_{\,\,\,\mu}$, which can be obtained from Equation~\ref{eq_str_ene_tens_2}. 
 The flux of $\mathcal{B}^{i}$ is $^\ast F^{ij}$, defined by Equation~\ref{duel_Faraday_tensor}. 
 The $D$, $T^{t}_{\,\,\,\mu}$, $^\ast F^{it}$ are then evolved through Equation~\ref{particle_conservation}, \ref{eq_ene-mom-cons} and \ref{B_evolution} advancing $dt_{n}$ in time. 
 The evolution of the electric filed $\pmb{\mathcal{E}}$ is implicit, hence cannot be evolved with other 
 conserved variables. 
 Still the {``}non-stiff part{''} ($N^{i}$ in Equation~\ref{E_evolution_numerical_2}) is a function only 
 of $\pmb{P}(t_{n})$, and is therefore calculated 
 using the primitive variables from the time step $t_{n}$.
 
 \item[(3)] We apply $U(t_{n+1})$ and the non-stiff part of $\pmb{\mathcal{E}}(t_{n+1})$ to the inversion scheme in 
  order to extract $\pmb{P}(t_{n+1})$ and the complete $\pmb{\mathcal{E}}(t_{n+1})$. 
  As discussed above (see Section \ref{2D+1_Scheme_subsec}), we first use $U(t_{n+1})$, except the electric field that 
  is taken from the previous time step, namely, $\pmb{\mathcal{E}}(t_{n})$. 
  We use the two equations in Equation~\ref{NR_eqations} and solve for the temporary primitive variables $u$, $v^{i}$ by applying 
  the 2D Newton-Raphson scheme. 
  We update $\pmb{\mathcal{E}}$ using Equation~\ref{E_evolution_numerical_1} with the temporary primitive 
  variables just obtained. 
  We return the temporary $\pmb{\mathcal{E}}$ and the temporary primitive variables back to the 2D Newton-Raphson
  scheme, and repeat this process until the primitive variables and $\pmb{\mathcal{E}}$ converge. 
  The converged primitive variables and $\pmb{\mathcal{E}}$ are now $\pmb{P}(t_{n+1})$ and $\pmb{\mathcal{E}}(t_{(n+1)})$.
 
 \item[(4)] The new time interval $dt$ is calculated considering also the diffusion time scale (see Section \ref{res_in_rHARM_chatper}), 
 together with $\pmb{P}(t_{n+1})$ (including $\pmb{\mathcal{E}}(t_{n+1})$ as primitive variables).
 We finally arrive at the evolutionary time step ($t_{n} \rightarrow t_{n+1}$). 

\end{enumerate}

Note that the actual time evolution in \HAR\ employs the simple first-second scheme (see equation 37 and 38 in \citealt{2013MNRAS.428...71B}), which is not shown in the flow chart in order to avoid complexity and confusion. 
But no structural changes are made at this point.   

%===============================================================================
\section{ B. 1D shock tube test for magnetic diffusivity}
\label{shock_tube_test}
Here we present test simulations of our implementation of magnetic diffusivity applying a classical shock tube setup \citep{2009JCoPh.228.6991D, 2013MNRAS.428...71B}. 
%The test simulations were executed in an 1D tube in Minkowski spacetime. 
%We have applied different levels of (constant) magnetic  diffusivity and
%compare the results to those in the above mentioned papers.  

\begin{figure*}
\centering
\includegraphics[width=3.5in]{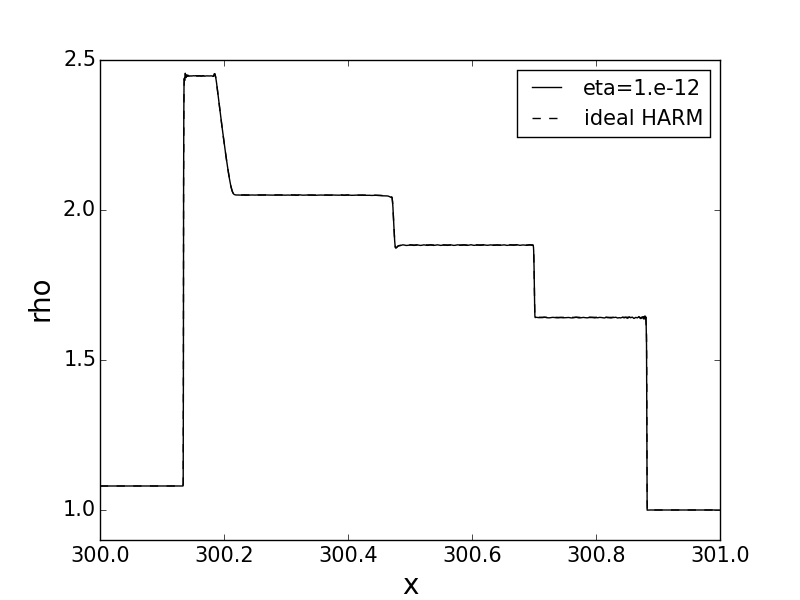}
\includegraphics[width=3.5in]{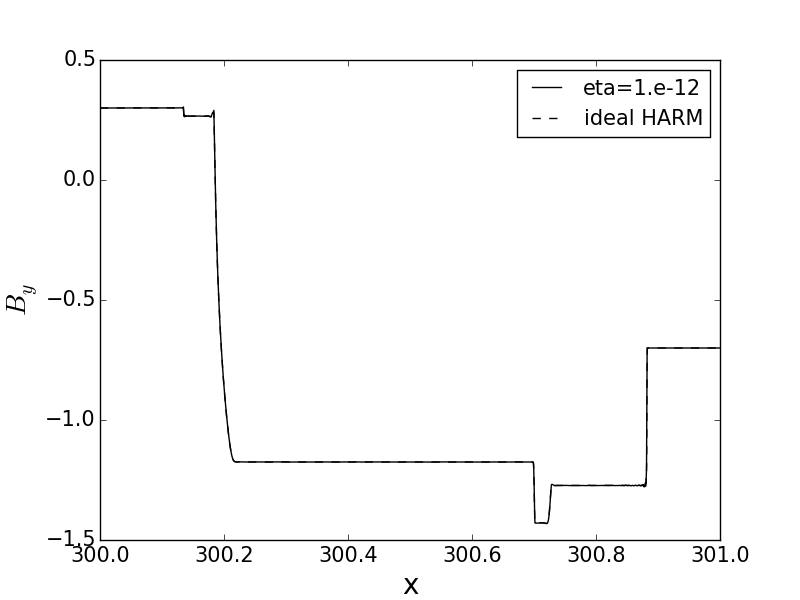}
\includegraphics[width=3.5in]{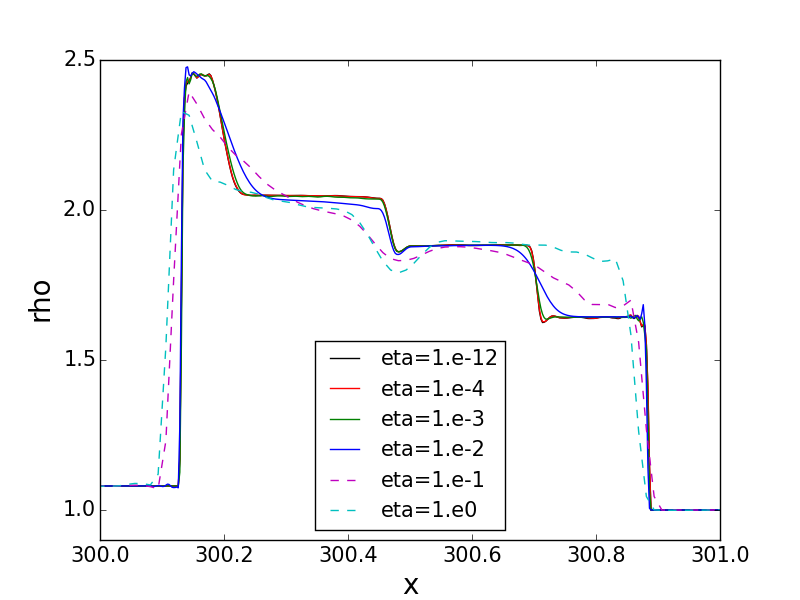}
\includegraphics[width=3.5in]{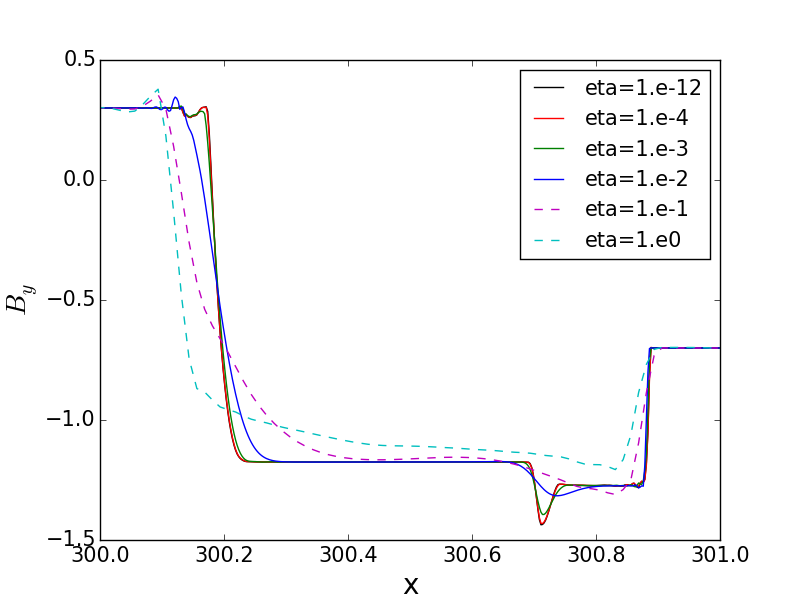}
\caption{Shock tube simulations. Density and vertical magnetic field at time $t=0.55t_{g}$. 
Above we show results of simulations with grid resolution $N=4500$ (where the curves of the two simulations match perfectly to each other),
below we show the results of simulations with $N=600$. 
In addition, the magenta and cyan dashed lines in the lower plots represents the simulations with resolution $N=120$. 
Although the actual computational domain is larger, only the range $x=[300 r_{g}, 301 r_{g}]$ is plotted in order 
to have a better comparison to the two reference papers. 
}
\label{shock_tube_results}
\end{figure*}

The shock tube simulation apply Minkowski space-time and Cartesian coordinates with equally spaced grid. 
The dimensional computational domain extends from $x_{0}$ to $x_{0} + 1.5 r_{g}$. 
In principal, $x_{0}$ can be set arbitrarily, here we chose $x_{0}=299.75 r_{g}$ and a computational domain $x=[299.75r_{g},301.25r_{g}]$.  
The initial conditions of primitive variables follow the setup in \citet{2009JCoPh.228.6991D} and \citet{2013MNRAS.428...71B} which are
\begin{equation} 
(\rho, p, v^{x}, v^{y}, v^{z}, B^{x}, B^{y}, B^{z}) = (1.08, 0.95, 0.4, 0.3, 0.2, 2.0, 0.3, 0.3) 
\label{shock_ini_l}
\end{equation}
for $x<300.5$ and
\begin{equation} 
(\rho, p, v^{x}, v^{y}, v^{z}, B^{x}, B^{y}, B^{z}) = (1.0, 1.0, -0.45, -0.2, 0.2, 2.0, -0.7, 0.5) 
\label{shock_ini_r}
\end{equation}
for $x>300.5$. 
The initial electric field is set to the ideal MHD value $E^{i} = - \epsilon^{ijk} B_{j} v_{k}$.
We apply Dirichlet boundary condition, where the primitive variables of both boundaries are fixed to the initial condition. 
The adiabatic index here is $\gamma=5/3$ following the test solutions from the literature.. 

We apply three different resolutions with $N=4500, 600, 120$ equidistant cells for different types of tests.
A first high resolution test was done with $N=4500$ and $\eta=10^{-12}$ in {\HAR} and then compared a similar
simulation applying the ideal HARM code. 
Both simulations match perfectly as shown in Figure \ref{shock_tube_results}. 
The curves also recover all features seen in \citet{2009JCoPh.228.6991D}. 
%This is a verification of the implementation in the ideal GRMHD ($\eta=0$) regime.

We then compare simulations runs with $\eta=10^{-12}$, $10^{-4}$, $10^{-3}$ and $10^{-2}$ with {\HAR} in
order to see the impact of diffusivity on the shock structure. 
Since the time stepping of the code becomes dominated by the diffusive time step for high diffusivity,
a resolution $N=4500$ cannot be reached.
Thus, for this set of tests we choose $N=600$ grid cells as resolution. 
The results are shown in the lower plots in Figure \ref{shock_tube_results} by solid lines. 
Due to the lower resolution, hence the larger numerical diffusivity, the discontinuities at the shock wave front 
for the $\eta=10^{-12}$ curve are broader than those for the high resolution plots.
Moreover, the $\eta=10^{-4}$ curve does not differ much from the $\eta=10^{-12}$ simulation,
essentially indicating a numerical diffusivity of similar order in this setup. 
Nevertheless, the two curves representing $\eta=10^{-3}$ and $\eta=10^{-2}$ nicely agree with those in \citet{2009JCoPh.228.6991D} and \citet{2013MNRAS.428...71B}. 

For simulations with $\eta > 10^{-2}$ we choose a resolution with $N=120$ cells for two reasons. 
At one hand, we recognized that some cell-scale oscillations that appear on the edge of the shock propagation 
(visible in Figure \ref{shock_tube_results} for $\eta>10^{-2}$) grow stronger and finally disturb the 
evolution of the shock propagation. 
A lower resolution can dissipate these oscillations (see below for a discussion). 
On the other hand, also the time costs of such diffusivity level also require a lower resolution.
However, note that the numerical diffusivity for this lower resolution is still below the physical diffusivity.
Our results for high diffusivity are shown by the dashed lines in the lower plots in Figure \ref{shock_tube_results}. 
The shape of these curves implies that the evolutions are still dominated by the physical diffusivity.
However, the discrepancy between these two simulations and those from the literature 
\citep{2009JCoPh.228.6991D, 2013MNRAS.428...71B} is obvious. 

We believe that instability appearing at the shock front mentioned above results from the shock capturing abilities
we use in \HAR. 
We find that this instability strongly depends on how the derivatives are calculated in the non-stiff term in Equation~\ref{E_evolution_numerical_2}. 
We have tried various limiters, such as monotonized central, van Leer and minmod slope limiter.
Different slope limiters always return slightly different results, but the problem could not be fixed by simply 
changing the slope limiter. 
Note also, that {\HAR} uses a simple first-second scheme for time evolution instead of the IMEX scheme applied in  \citet{2009MNRAS.394.1727P}, which was also employed in \citet{2013MNRAS.428...71B}.
This might add to the inaccuracy of the code in high-$\eta$ regime as well. 
However, since the magnetic diffusivity values we apply in our accretion-ejection setup will be always below 
$\eta = 10^{-2}$, we decided that - at this point in time - not to go deeper into this problem.

%===================================================================================================
\section{ C. Numerical diffusivity}
\label{num_diff_chapter}
Having implemented a physical magnetic diffusivity, we are able to measure the numerical
diffusivity of \HAR.
In order to do so, we have run the setup of section 6 for an extended parameter range concerning numerical resolution
and physical magnetic diffusivity $\eta$.
For a box size of $(\Delta r \times r \Delta \theta) = (1.0\times 1.0)$, located at $r=300$ we applied numerical grids of
$(16\times 16)$, $(32\times32)$, $(64\times64)$, $(128\times128)$, and $(256\times256)$.
Depending on resolution, we applied physical magnetic diffusivities between $\eta=10^{-8}$ and $10^{-3}$.

\begin{figure*}
\centering
\includegraphics[height=2.6in]{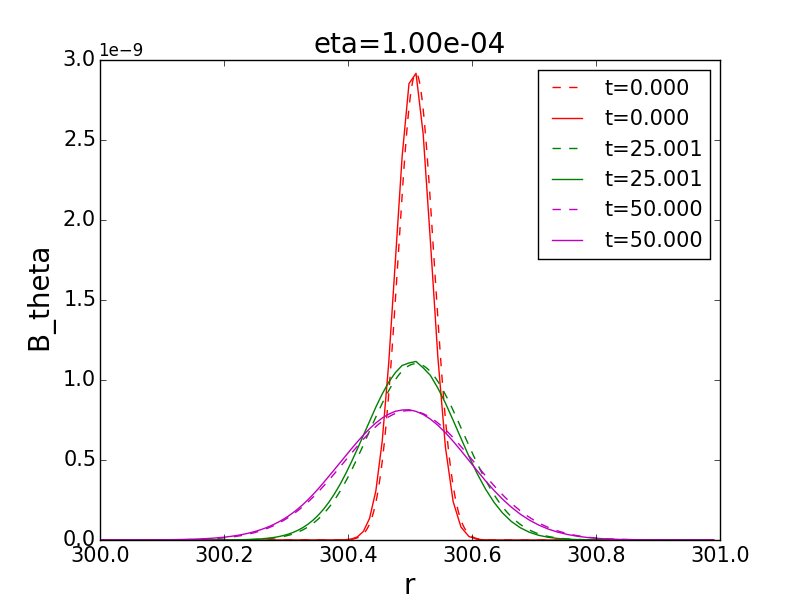}
\includegraphics[height=2.6in]{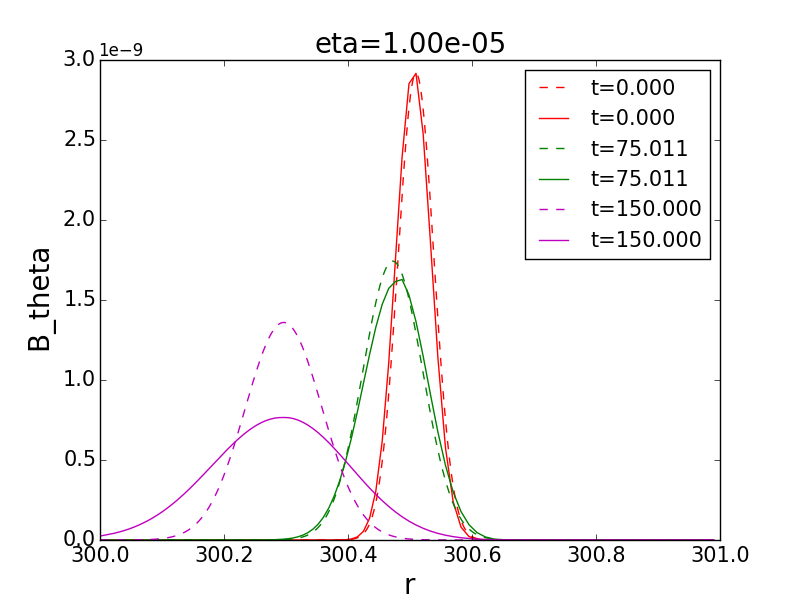}
\includegraphics[height=2.6in]{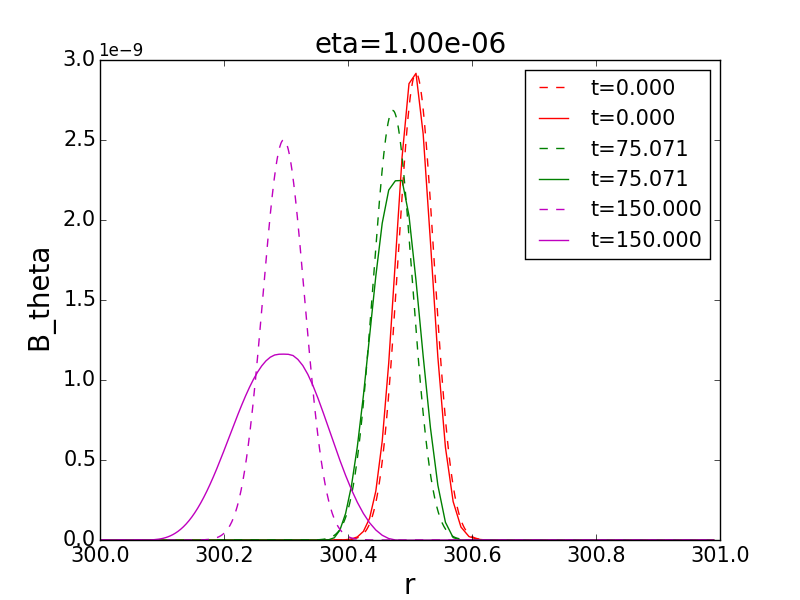}
\includegraphics[height=2.6in]{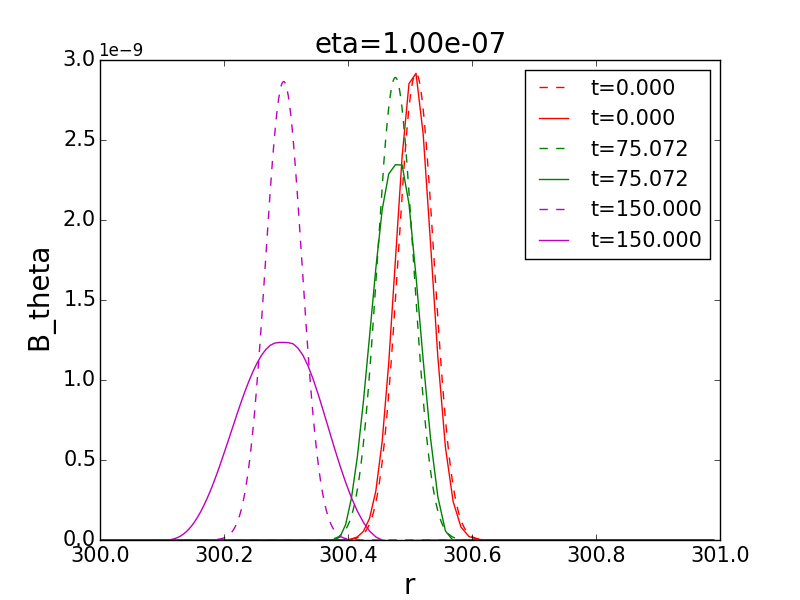}
\includegraphics[height=2.6in]{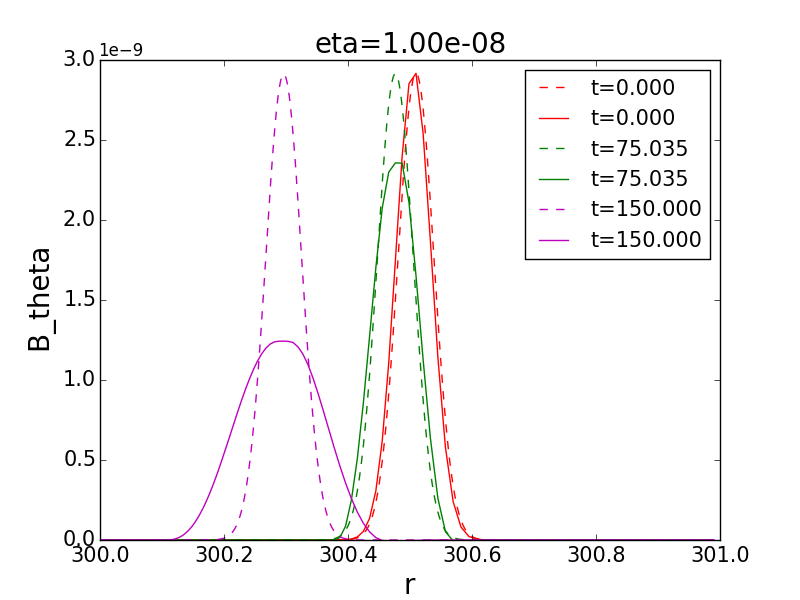}
\caption{Measure of numerical magnetic diffusivity.
The simulations shown apply a numerical resolution of $(128\times128)$,
and a physical magnetic diffusivity $\eta = 10^{-8}, 10^{-7}, 10^{-6}, 10^{-5}, 10^{-4}$ 
(from bottom right to top left, see figure titles).
Dashed curves show the analytic solution of the diffusion equation for the physical
magnetic diffusivity (as in Section \ref{diff_box_chapter}).
Solid curves show the result of the numerical simulation for the same time steps.
% in Boyer-Lindquist coordinates.
% This is simulation: /HARM-QQ-NEW/Standard/Run09_Christos_paras/Sim03
}
\label{fig_num_diff}
\end{figure*}

Examples of our runs for resolution  $(128 \times 128)$ are shown in Figure \ref{fig_num_diff}.
and for a physical magnetic diffusivity $\eta = 10^{-8}, 10^{-7}, 10^{-6}, 10^{-5}, 10^{-4}$ (from bottom right to top left).
The dashed curves show the analytic solution of the diffusion equation for the physical
magnetic diffusivity, while  the solid curves show the result of the numerical simulation for the same 
time steps.
Note that we did not correct for the mass infall (see discussion in section 6). This results in the slight acceleration of 
the material towards the black hole, and the average infall velocity of 
$v_{\rm fall} \equiv \Delta r \Delta t \simeq 0.2/150 = 1.3\times 10^{-3}$ (in code units).

For $\eta = 10^{-4}$, the numerical simulation follows the analytic solution, indicating that the decay of the magnetic 
field obeys the physical magnetic diffusivity. 
We run this simulation only till $t=50$, which is about the diffusive time scale for $\eta = 10^{-4}$ for the analytic solution.
For $\eta = 10^{-5}$, numerical diffusivity seems to contribute, in particular for the later time steps.
For $\eta = 10^{-6}$ numerical diffusivity is dominating, as the analytic solution would only
slightly decay within time frame applied.
For $\eta = 10^{-7}$, the numerical evolution of the magnetic field is practically identical to the simulation results of $\eta = 10^{-6}$,
telling that the systems now evolves only under numerical diffusivity.
Clearly, the physical diffusivity prescribed is so small that it plays no role, and the system evolves only
under numerical resistivity.
We conclude that for the given setup and resolution, the numerical diffusivity is of the
order of  $\eta_{\rm num} \sim 10^{-5}$.

A similar study with resolution $(256\times 256)$ shows that in this case the numerical diffusivity is of the
order $\eta_{\rm num} \sim 10^{-6}$,
while for $(64 \times 64)$ the numerical diffusivity is $\eta_{\rm num} \sim 10^{-4}$.

As an alternative study, we investigated for a fixed (physical) magnetic diffusivity of $\eta = 10^{-5}$
(or also $\eta = 10^{-4}$) and a range of grid resolutions. 
In agreement with the previous study we find that the numerical magnetic diffusivity equalize the physical 
magnetic diffusivity for a certain resolution and dominates the physical diffusivity for lower resolution.

The exact number values clearly depend on the numerical setup, but in order to evolve physical diffusion processes
with \HAR\ with a physical diffusivity of $\eta = 10^{-4} ... 10^{-2}$ for a grid resolution of $\leq 10^{-2}$ 
is needed.
Also, for a physical magnetic diffusivity varying in space (for example a disk magnetic diffusivity) the numerical
diffusivity serves as a {"}floor{"} value - with a value depending on resolution.

%---------------------------------------------------------------
%\begin{bibliography}{}
%\bibliographystyle{apj}
%\bibliography{paper I.bib}
%\end{bibliography}

% THE REFERENCES ARE HERE
%\bibliographystyle{apj}
%\bibliography{main}

%\clearpage

\bibliographystyle{apj}

% THE REFERENCES ARE HERE
%\bibliography{jet_paper_new}

% THE REFERENCES ARE HERE

%\input{paper_nezami.bbl}
%---------------------------------------------------------------
\end{document}